\documentclass[pra,twocolumn,aps,superscriptaddress,floatfix]{revtex4-1}
\usepackage[utf8]{inputenc}
\usepackage{dsfont}
\usepackage{amsmath}
\usepackage{amssymb}
\usepackage{graphicx}
\usepackage{color}
\usepackage{float}
\usepackage[bookmarks=false,linkcolor=blue,urlcolor=blue,colorlinks,citecolor=blue]{hyperref}

\newcommand{\ket}[1]{|#1\rangle}
\newcommand{\bra}[1]{\langle #1|}
\newcommand{\ketbra}[2]{|#1\rangle\langle #2|}
\newcommand{\braket}[2]{\langle #1|#2\rangle}
\newcommand{\matelem}[3]{\langle #1|#2|#3\rangle}
\newcommand{\expect}[2]{\langle #1|#2|#1\rangle}
\newcommand{\average}[1]{\langle #1\rangle}
\newcommand{\env}{\mathrm{env}}

\newcommand{\notop}{{\phantom{\dag}}}
\newcommand{\nodag}{{\phantom{\dag}}}

\newcommand{\Tr}{\mathrm{Tr_{env}}}
\newcommand{\nF}{n_\mathrm{F}}
\newcommand{\nB}{n_\mathrm{B}}
\newcommand{\kB}{k_\mathrm{B}}
\newcommand{\rrho}{\boldsymbol{\rho}}

\newcommand{\rrr}{\mathbf{r}}

\newcommand{\abs}[1]{\left\vert #1 \right\vert }

\usepackage{bm}
\usepackage{bbm}

\begin{document}
\title{Fidelity and visibility loss in Majorana qubits\\ by entanglement with environmental modes}

\author{Morten~I.~K. Munk,$^1$ Reinhold Egger,$^2$ and Karsten Flensberg}
\affiliation{Center for Quantum Devices, Niels Bohr Institute, University of Copenhagen, 2100 Copenhagen, Denmark,\\
$^2$Institut f\"ur Theoretische Physik, Heinrich-Heine-Universit\"at, 40225  D\"usseldorf, Germany}

\begin{abstract}
We study the dynamics and readout of topological qubits encoded by zero-energy Majorana bound states in a topological superconductor. We take into account bosonic modes due to the electromagnetic environment which couple the Majorana manifold to above-gap continuum quasi-particles. This coupling causes the degenerate ground state of the topological superconductor to be dressed in a polaron-like manner by  quasi-particle states and bosons, and the system to become gapless. Topological protection and hence full coherence is only maintained if the qubit is operated and read out within the low-energy spectrum of the dressed states. We discuss reduction of fidelity and/or visibility if this condition is violated by a quantum-dot readout that couples to the bare (undressed) Majorana modes. For a projective measurement of the bare Majorana basis, we formulate a Bloch-Redfield approach that is valid for weak Majorana-environment coupling and takes into account constraints imposed by fermion-number-parity conservation. Within the Markovian approximation, our results essentially confirm earlier theories of finite-temperature decoherence based on Fermi's golden rule. However, the full non-Markovian dynamics reveals, in addition, the fidelity reduction by a projective measurement. Using a spinless nanowire model with $p$-wave pairing, we provide quantitative results characterizing these effects.
\end{abstract}
\date{\today}
\maketitle

\section{Introduction}\label{sec1}

Currently there is a large interest in topological phases with defects that can nonlocally store quantum information and thus possibly offer avenues to topologically protected quantum information processing \cite{Kitaev2003,NayakReview}. One such example is a topological superconductor (TS) wire which  supports Majorana bound states (MBSs) at its ends \cite{Kitaev2001}.  Because it takes two MBSs to form a fermionic level, the occupancy of this level is stored nonlocally when the MBSs are spatially well separated. As a consequence, under ideal conditions, the quantum information can neither be retrieved by a local measurement nor be destroyed by local noise sources. The search for MBSs has intensified since the appearance of theoretical proposals in hybrid systems made of superconductors and semiconductors \cite{Lutchyn2010,Oreg2010,BeenakkerReview,AliceaReview,AguadoReview,LeijnseReview,LutchynReview} or topological insulators \cite{Fu2008}.
Several tunneling spectroscopy experiments have already been published and appear to be consistent with the existence of MBSs  \cite{Mourik2012,Deng2016,Albrecht2016,Suominen2017,Nichele2017,Zhang2018}.

The prospect of robust MBS realizations in solid state systems has spurred many proposals for Majorana based qubits \cite{Hyart2013,Aasen2016,Plugge2017,Karzig2017,Manousakis2017} and for error correction schemes \cite{Terhal2012,Landau2016,Brell2014,Vijay2015,Plugge2016,Litinski2017,Litinski2018a}. The latter can correct errors due to, e.g., quasi-particle poisoning caused by spurious fermionic excitations. Majorana based architectures do not have a universal set of topologically protected gates and are limited to Clifford gates only.  The above-mentioned schemes must therefore be augmented by non-protected gates in order to achieve universal quantum computation \cite{Bravyi2006,Flensberg2011,Hyart2013,Plugge2016,Karzig2016,Karzig2017}. More complex anyon excitations, e.g., Fibonacci anyons, would allow to implement a universal set of topologically protected gates \cite{NayakReview}. However, such systems are still far from experimental realization.

\begin{figure}[t]
\includegraphics[width=0.43\textwidth]{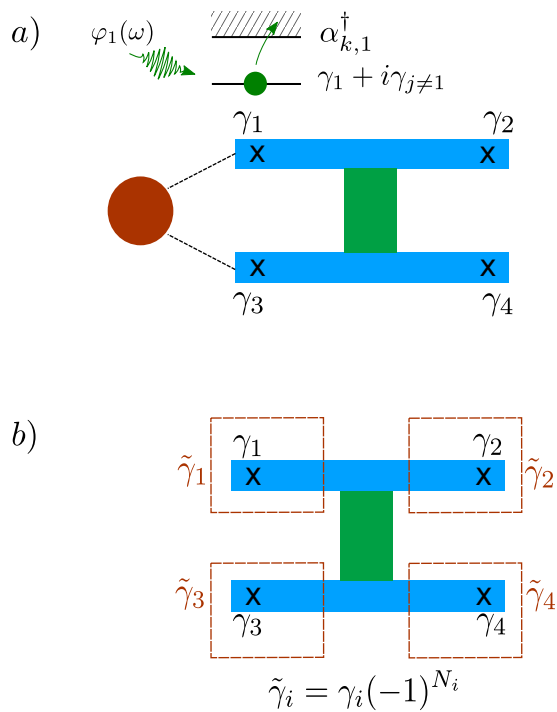}
\caption{Basic setup and two types of qubit readouts with $M=4$ MBSs. Each MBS is coupled to independent local charge fluctuations. The blue horizontal bars represent TS wires which are connected by a conventional superconductor bridge (green vertical bar). The basic mechanism for decoherence is also illustrated for bosons representing voltage fluctuations, $\varphi_{1}(\omega)$, which cause coupling between the MBS sector and the gapped quasi-particle sector. (a): \textit{Bare Majorana readout}. With a tunnel-coupled quantum dot (red circle), one can read out $i\gamma_1\gamma_3$ \cite{Plugge2017,Karzig2017,Flensberg2011}. (b): \textit{Total fermion-parity encoding} where the qubit information is stored in the combined parity of the MBS plus the local quasi-particle continua, mathematically represented by the modified Majorana operators $\tilde\gamma_i$. In principle, this qubit is immune to local charge fluctuations. However, manipulation and readout by, e.g., control of total charge in each arm is practically very difficult. In this paper, we focus on case (a).}
\label{fig1}
\end{figure}

Majorana qubits are often argued to have long coherence times because of the underlying topological protection. The usual reasoning is that because no local operator can split the topological ground-state degeneracy, the quantum information is protected against local perturbations as long as the MBSs are non-overlapping. For finite MBS overlap, the protection of the ground-state degeneracy is lifted and protection is lost. This case has recently been analyzed in Ref.~\cite{knapp2018}. Even when direct MBS overlaps are negligibly small, as will be assumed in our work, boson-mediated couplings of MBSs to above-gap quasi-particles cause a coherence decay at finite temperatures
\cite{Goldstein2011,Cheng2012,Rainis2012,Schmidt2012,Yang2014,Lai2018,Song2018,Li2018,Nag2018,Aseev2018}. The bosonic modes could represent, for instance, phonons or fluctuating charge degrees of freedom.
This finite-temperature decoherence mechanism follows from a Markovian approximation, i.e., by assuming a negligible memory time of the environment. Available estimates of the corresponding decoherence rate, $\Gamma$, obtained by assuming either uniform \cite{Rainis2012,Schmidt2012} or non-uniform \cite{Aseev2018} gate voltage fluctuations, suggest that coherent qubit operation may be hard to achieve on above-microsecond time scales even though the rate is exponentially small, $\Gamma\propto T\exp(-\Delta/\kB T)$, with the TS gap $\Delta$ and temperature $T$, see, e.g., Ref.~\cite{Schmidt2012}.  Recent work has also studied the fault tolerance threshold for Majorana qubits in a similar setting \cite{Knapp2018b}.

One of our goals is to address what happens in the non-Markovian case, both at $T=0$ and finite $T$. We consider a specific encoding, which we denote a \textit{bare-Majorana qubit} (Fig.~\ref{fig1}a), where the qubit space is addressed by quantum dots that couple to the \emph{uncoupled} and \emph{undressed} MBSs, as schematically illustrated in Fig.~\ref{fig1}(a). This setup has, for example, been proposed for measurement-based manipulations of a topological quantum computer \cite{Karzig2017,Plugge2017}. In such setups, MBSs couple both to readout devices, e.g., the dot in Fig.~\ref{fig1}(a), and to bosonic and quasi-particle environments.

In order to analyze the non-Markovian dynamics of the bare-Majorana qubit, we develop and apply a modified Bloch-Redfield master equation approach \cite{Weiss2010} which is valid for weak Majorana-environment coupling. Moreover, we also analyze the read-out protocol by using perturbation theory. Employing the Bloch-Redfield equations, we investigate the dynamics of bare-Majorana qubits formed from $M$ non-overlapping MBSs  in the presence of local (quantum) charge fluctuations. The interaction between these fluctuations and the MBSs implies that eigenstates of the entire system exhibit entanglement between the MBS sector and the environment. (The latter is formed by quasi-particles and the bosonic modes describing charge fluctuations.) Because of this entanglement, topological protection is only preserved for a combined fermion parity degree of freedom (combined continua and MBSs), cf.~Fig.~\ref{fig1}(b) and Sec.~\ref{sec2b} below, and not for the isolated MBS manifold.

The entanglement of MBSs and environmental modes can physically be understood as result of virtual (off-shell) processes. As a consequence, all coherences of the bare (undressed) Majorana system will be reduced by a factor $1/(1+\eta)$ at long times. For the ground state, the reduction factor corresponds to the squared overlap between the true polaron-like ground state and the bare ground state and it is relevant for a \textit{projective} measurement of the state of the bare MBS. We discuss qualitatively how  this theory can be adapted to the case of slow turn-on of the measurement circuit. Even though the system is gapless, it should still be possible to minimize the effect of quasi-particle generation if one effectively reads out the entangled (polaron-like) states by carefully timing the read-out device. Moreover, by engineering of the electromagnetic environment it would be possible to improve on the adiabaticity condition, for example by creating a gapped or reduced low-energy environment spectrum.

The minimal setup with $M=4$ is illustrated in Fig.~\ref{fig1}(a), where the MBSs forming the qubit are individually coupled to independent local charge fluctuations. The qubit state can then be read out, for example, by coupling a MBS pair to a nearby quantum dot \cite{Flensberg2011,Plugge2017,Karzig2017}. We emphasize that the four edge regions of the qubit are not coupled in our analysis. Therefore the effects we consider all result from a system which in principle has topological protection. Effects beyond this model, for example finite overlap of the MBSs or the quasi-particle states, comes on top of our analysis. The topological protection here means that if one operates and reads out the qubit in a total-parity basis, coherence is fully maintained. This could for example be done by a charge readout after disconnecting sections of the Majorana system as suggested by Aasen \textit{et al.}\cite{Aasen2016}.

The paper is organized as follows. In Sec.~\ref{sec2}, we define the bare-Majorana qubit encoding and discuss an alternative fermion-parity qubit encoding \cite{Akhmerov2010}, which would be free from decoherence but seems difficult to realize in practice. In Sec.~\ref{sec3}, we then explain the physics of the Majorana-environment coupling. In Sec.~\ref{sec:readout}, we give a simple physical argument for the reduction of fidelity based on first-order perturbation theory. In Sec.~\ref{sec:readoutsudden}, we consider projective readout of the bare MBS. For this purpose, we develop in Sec.~\ref{sec5} a Bloch-Redfield master equation approach  for studying the dynamics of a bare-Majorana qubit. The Markovian limit is discussed in Sec.~\ref{sec5b}, followed by a study of non-Markovian effects Sec.~\ref{sec5c}. In Sec.~\ref{sec6}, we apply this theory to a specific case where MBSs and quasi-particles originate from a spinless TS wire with $p$-wave pairing symmetry \cite{AliceaReview}. In Sec.~\ref{sec6a}, we address the finite-$T$ case, and in Sec.~\ref{sec6b} our $T=0$ results are presented.  The paper closes in Sec.~\ref{sec7} with a summary and concluding remarks.  Technical details have been delegated to several Appendices.

\section{Qubit readout}
\label{sec2}

In this section, we discuss how the quantum information is addressed in Majorana-based qubits and we distinguish between two different principles. The first relies directly on the zero modes such that coupling to MBSs, for example, via quantum dots \cite{Karzig2017,Plugge2017} is used to read out or initialize the qubit state. We refer to this setup as a \emph{Bare-Majorana qubit}. The second method represents a \emph{Total parity qubit}. The latter requires measurements of total parities which in turn necessitate tunable Josephson junctions as, e.g., in the proposal of Ref.~\cite{Aasen2016}. The difficulty is therefore in choosing the right time scales for switching on and off the coupling between the various segments of the qubit, a problem  analyzed in Ref.~\cite{Hell2016}. In this paper, we investigate the decoherence dynamics of a bare-Majorana qubit.

\subsection{Bare-Majorana qubit}
\label{sec:MBSqubit}

Consider a Majorana island as the one depicted in Fig.~\ref{fig1}(a), where MBSs correspond to self-adjoint Majorana operators, $\gamma^\nodag_j=\gamma_j^\dagger$, with $j=1,\ldots,M$ and  anticommutation relations $\{\gamma_j,\gamma_{j'}\}=2\delta_{jj'}$.
For more detailed device layouts and measurement schemes, see Refs.~\cite{Karzig2017,Plugge2017}. A quantum dot is tunnel coupled to two MBSs for the purpose of reading out the joint MBS parity. The qubit with attached read-out device is described by the Hamiltonian
\begin{equation}\label{HdotcoupleMBS}
\begin{aligned}
  H=&   \, \varepsilon_d^\nodag c^\dag_dc_d^\nodag +\sum_{i} \left(t_i^*c_d^\dag \Psi(r_i)+\mathrm{h.c.}\right) \\
  &\quad+ H_\mathrm{qubit}+E_C(N_\mathrm{qubit}-N_g)^2,
\end{aligned}
\end{equation}
where $c_d$ is the fermionic dot-level annihilation operator, $\Psi(r_i)$ is the electron operator in the TS taken at the position of the tunnel coupling to MBS $i$, $t_i$ is the corresponding tunneling amplitude,  $H_\mathrm{qubit}$ describes the qubit with its coupling to other environments, see Sec.~\ref{sec3}, $N_\mathrm{qubit}$ is the qubit total electron-number operator, $N_g$ is a dimensionless gate potential, and finally, $E_C$ is the charging energy of the Majorana island. If the readout is done measuring the quantum charge by a charge sensor, the readout device is effectively distinguishing the derivative of the energy:
\begin{equation}\label{nderivationenergygeneral}
  \langle n_d\rangle =\frac{d\langle H\rangle}{d\varepsilon_d} .
\end{equation}
Similarly, if instead the capacitance of a circuit is measured, the readout device effectively reads out the second derivatives $d^2 \langle H\rangle/d\varepsilon^2_d$ \cite{Karzig2017}.

We assume the system is tuned so that the charge configuration, $(n_d,N_\mathrm{qubit})$, is near $(0,0)$ and $(1,-1)$. In this case, the Hamiltonian \eqref{HdotcoupleMBS} becomes (up to a constant)
\begin{equation}\label{HdotcoupleMBS2}
  H=\tilde{\varepsilon}_d^\nodag c^\dag_dc_d^\nodag +\sum_{i} \left(t_i^*c_d^\dag \Psi(r_i)+\mathrm{h.c.}\right)+ H_\mathrm{qubit},
\end{equation}
where $\tilde{\varepsilon}_d={\varepsilon}_d+E_C(1+2N_g)$.

When projecting the qubit to its low-energy subspace, we replace the electron operator by the respective Majorana operator, $\Psi(r_i)\approx a_i\gamma_i$, where $a_i$ is the value of the electron component of the MBS wave function at $r_i$.   We then include the $a_i$ in the definition of the tunnel couplings $t_i$.

{We can only solve for the energy in the case where $H_\mathrm{qubit}=0$, i.e., for the ideal situation without environmental degrees of freedom.
In this case, assuming that the quantum dot is tunnel-coupled to $\gamma_1$ and $\gamma_3$ only,  the energies $E_{s=\pm}$ of the split ground-state manifold are given by}
\begin{equation}\label{Energygeneralnoqubit}
  { E_s =\frac{\tilde{\varepsilon}_d}2- \sqrt{\left(\frac{\tilde{\varepsilon}_d}2\right)^2+|t_1|^2+|t_3|^2-2s \mathrm{Im}[t_1t_3^*]},}
\end{equation}
{where the combined parity $s=(-1)^{(i\gamma_1\gamma_3 - 1)/2 + n_d}$ is a good quantum number.}

{When the qubit Hamiltonian is non-zero, we need to study the tunneling Hamiltonian perturbatively. For small $t_i$, and assuming that $N_g$ is tuned to a value where the quantum dot is empty ($c_d^\dag c_d^\nodag=0$) without tunneling, second-order perturbation theory gives the effective Hamiltonian} \cite{Karzig2017}
\begin{equation}\label{HdotcoupleMBS2ndorder}
 {H^{(2)}=\sum_{i,j} \frac{2i\gamma_i^\nodag\gamma_j^\nodag\mathrm{Im}[t_i^\nodag t_j^*] -|t_i|^2-|t_j|^2}{2\tilde{\varepsilon}_d}  + H_\mathrm{qubit}.}
\end{equation}
{The expression \eqref{HdotcoupleMBS2ndorder} is perturbative in the tunneling coupling and valid away from the charge-degeneracy point. Evidently, in this regime, when the dot is only coupled to $\gamma_1$ and $\gamma_3$, then $i\gamma_1\gamma_3 = \pm 1$ is a good quantum number.}

To summarize, in this subsection we have discussed various readout schemes of Majorana qubits. When reading out the parity of two MBSs using a quantum dot, the read-out device couples (for $M=4$) to the Pauli operators
\begin{equation}\label{paulidefs}
  \sigma_x=i\gamma_1\gamma_2,\quad  \sigma_y=i\gamma_2\gamma_3,\quad   \sigma_z=i\gamma_1\gamma_3.
\end{equation}
However, when the Majorana qubit is coupled to other degrees of freedom, the qubit as defined in Eq.~\eqref{paulidefs} is no longer well-defined (because the Pauli operators $\sigma_i$ do not necessarily commute with $H_\mathrm{qubit}$) and one needs to discuss the influence on the readout fidelity and/or readout visibility. This is the main purpose of this paper. In Sec.~\ref{sec3}, we set up our model for the qubit \eqref{paulidefs} in the presence of environmental modes. In the subsequent sections, we then study the influence of qubit-environment entanglement on the qubit dynamics.

\subsection{Total-parity qubit }
\label{sec2b}

As an alternative to the bare-Majorana readout discussed above, one can define a set of Pauli operators based on the total number parity of each region which is fully protected against decoherence. This approach was pointed out by Akhmerov \cite{Akhmerov2010} who showed that topological protection is maintained as long as different MBSs do not interact directly or via continuum states. Instead of  Eq.~\eqref{paulidefs}, one defines Pauli operators by taking into account the total number of fermions in each spatial region,
\begin{eqnarray}\label{taupaulidefs}
  \tilde\sigma_x &=&\sigma_x \left(-1\right)^{N_1+N_2},\quad
\tilde\sigma_y=\sigma_y \left(-1\right)^{N_2+N_3},\\ \nonumber
  \tilde\sigma_z&=&\sigma_z \left(-1\right)^{N_1+N_3},\quad N_i=\sum_k\alpha_{k,i}^\dag\alpha_{k,i}^\nodag,
\end{eqnarray}
where the operator $N_i$ counts the number of above-gap quasi-particles in the respective region, cf.~Sec.~\ref{sec3}.
 It is easy to check that the  $\tilde\sigma_{x,y,z}$ satisfy the Pauli algebra, e.g.,
\begin{equation}\label{taupaulialg}
  \tilde\sigma_x\tilde{\sigma}_y =i {\sigma_z \left(-1\right)^{N_1+2N_2+N_3}}=i\tilde\sigma_z.
\end{equation}
In addition, all $\tilde\sigma$ matrices commute with the full Hamiltonian $H$ (including the environmental degrees of freedom), which in turn conserves all parities
  associated with pairs of regions,
\begin{equation}\label{Nijdef}
  P_{ij}= \left[(i\gamma_i\gamma_j-1)/2+N_i+N_j\right]~{\rm mod}~2.
  \end{equation}
Another way to understand this fact is to verify that the modified Majorana operators
\begin{equation}\label{tildegammedeff}
\tilde\gamma_j=\gamma_j (-1)^{N_j}
\end{equation}
commute with  $H$. We refer to Fig.~\ref{fig1}(b) for an illustration of the total-parity Majorana operators.

The new Pauli operators \eqref{taupaulidefs} represent quantum information that is topologically protected and can only be corrupted by finite size effects, causing MBS wave function overlap or transfer of quasi-particles between different MBS regions. However, in practice this protection can only be employed if one is able to manipulate and read out in this basis. This could in principle be performed by using the charging energy to fuse two MBSs \cite{Hyart2013, Aasen2016} which would require tunable Josephson junctions that can be tuned to the closed regime, thereby limiting the allowed time scales \cite{Hell2016}. However, the coupling to environmental bosons imposes further restrictions because of the absence of a gap.

\section{Coupling of Majorana states to environment}
\label{sec3}

We now describe a general model for studying how the dynamics of a Majorana-based qubit is affected by the coupling between  MBSs and environmental degrees of freedom.
By environmental modes, we here mean above-gap TS quasi-particles  and bosonic modes corresponding to electric potential fluctuations.
 Let us begin with the unperturbed superconducting system in the absence of charge fluctuations. It is governed by the Hamiltonian
\begin{equation}\label{H0BCS}
  H_0=\frac12\int d\rrr\,\Psi^\dag(\rrr)\mathcal{H}_\mathrm{BdG}\Psi(\rrr),
\end{equation}
where we define 4-spinors,
\begin{equation}\label{nambu}
\Psi(\rrr)=\left(\Psi_\uparrow^\nodag(\rrr),\Psi_\downarrow^\nodag(\rrr),\Psi_\downarrow^\dag(\rrr),-\Psi_\uparrow^\dag(\rrr)\right)^T,
\end{equation}
with the electron annihilation operator
$\Psi^\nodag_\sigma(\rrr)$ for spin $\sigma=\uparrow,\downarrow$ and position $\rrr$.  We use
Pauli matrices $\tau_{x,y,z}$ in Nambu (particle-hole) space. The Bogoliubov-de Gennes (BdG) Hamiltonian appearing in Eq.~\eqref{H0BCS} corresponds to the Nambu matrix
\begin{equation}\label{BdG}
  \mathcal{H}_\mathrm{BdG}=\left(
\begin{array}{cc} \mathcal{H}_0 & \Delta \\
\Delta^\dag & -\mathcal{T} \mathcal{H}_0 \mathcal{T}^{-1} \\
\end{array} \right),
\end{equation}
where $\mathcal{H}_0$ is the spinful single-electron  Hamiltonian   in the absence of pairing (and, of course, without charge fluctuations), $\Delta$ is the pairing potential in BCS mean-field approximation, and $\mathcal{T}$ is the time-reversal operator. After diagonalizing the BdG Hamiltonian, the Hamiltonian \eqref{H0BCS} can be written in terms of BdG quasi-particle eigenmodes corresponding to a set of annihilation operators $\alpha_k$. The $\alpha_k$ operators describe fermionic eigenstates with energy  $E_k\ge \Delta$, where quantum numbers $k$ label different eigenmodes.  Consequently, Eq.~\eqref{H0BCS} takes the form
\begin{equation}\label{BdG2nd}
  H_0=\sum_k E_k^\notop \alpha_k^\dag\alpha_k^\nodag + \mathrm{constant}.
\end{equation}
In the topological phase, an even number $M$ of localized zero-energy MBSs can be present in addition. In particular, for 1D TS wires, MBSs
exist at each end of a topological wire segment.
 As the Majorana operators $\gamma_j$ describe zero-energy modes, they
 do not appear in $H_0$ and thus also commute with the unperturbed Hamiltonian, $[H_0,\gamma_j]=0$ \cite{BeenakkerReview,AliceaReview,AguadoReview,LeijnseReview,LutchynReview}.

We next note that $H_0$ implicitly includes the electric potential in the superconducting material. If this potential can change due to fluctuations mediated by other (bosonic) degrees of freedom, it must be included in the model. The full Hamiltonian, $H=H_{\rm qubit}$, is then given by
\begin{equation}\label{dHincluded}
  {H}={H}_0+{H}_\varphi+H_\mathrm{int},\quad H_\mathrm{int}=\int d\rrr\,\rho_e(\rrr) \varphi(\rrr),
\end{equation}
where $\varphi(\rrr)$ is an operator that describes the electric potential fluctuations caused by a set of bosonic modes. The potential fluctuations occur, in principle, on all length scales. For simplicity, we here focus on the most important components, namely the potential fluctuations with length scales of order the coherence length. Hence, we replace $\varphi(\rrr)$ by $M$ independent fluctuating potentials, $\varphi_j$, one for each region $j=1,\ldots,M$. The bare dynamics of these fluctuations are governed by a non-interacting bosonic Hamiltonian, $H_\varphi$. In principle, one could also include fields describing fluctuations of the magnetic field, but for simplicity we focus on electrical fluctuations below.

Expressing the electron density $\rho_e(\rrr)$ in Eq.~\eqref{dHincluded} in terms of BdG quasi-particle operators, we get two contributions, $H_\mathrm{int}=H_1+H_2$, with
\begin{equation}\label{H1def}
  H_1=\sum_{j=1}^M \gamma_j \Gamma_j \varphi_j ,
 \quad \Gamma_j =\sum_k \left(W_{k,j}^\notop\alpha^\dag_{k,j}-W_{k,j}^\ast\alpha_{k,j}^\notop\right),
\end{equation}
and
\begin{equation}\label{H2def}
  H_2= \sum_{k,k',j}\left(V_{kk'j}^{(1)}
  \alpha^\dag_{k,j}\alpha^\notop_{k',j}
  +V_{kk'j}^{(2)}\alpha^\dag_{k,j}\alpha^\dag_{k',j} \right)\varphi_j + {\rm h.c.}
\end{equation}
We here define the $W$ matrix elements as
 \begin{equation}\label{Wkirdef}
  W_{k,j}=  \langle k,j|\tau_z | \mathrm{MBS},j\rangle,
\end{equation}
where $|k,j\rangle$  ($|\mathrm{MBS},j\rangle$) denotes a BdG quasi-particle (MBS) spinor wave function in the $j^\mathrm{th}$ region.
For concrete results,
one has to consider a specific model for the TS nanowire. In Sec.~\ref{sec6}, see also App.~\ref{App::WandJ}, we discuss the matrix elements
\eqref{Wkirdef} for a semi-infinite spinless TS wire model with $p$-wave pairing.

To recapitulate, the above model Hamiltonian describes coupling between a TS and bosonic potential fluctuations. To emphasize the important physics studied in this paper, we have made the following key simplifications: (\textit{i}) All MBSs are treated as non-overlapping zero-energy states. (\textit{ii}) Quasi-particle modes described by the fermionic operators $\alpha_{k,j}$ are assumed to have no significant support in spatial regions where other MBSs reside, and hence no MBS-MBS interactions are mediated through continuum states either. (\textit{iii}) The charge density $\rho_e(\rrr)$ in the region near the $j^\mathrm{th}$ MBS couples to an operator $\varphi_j$ describing the long wave length component of the field in that region.  Given the typically small size of these regions, we neglect the spatial dependence of  $\varphi_j$. (\textit{iv}) We assume that different $\varphi_j$ operators are uncorrelated, i.e., each MBS is independently coupled to its own fluctuating electric field. (\textit{v})  The $V^{(1,2)}$ matrix elements in Eq.~\eqref{H2def} are not important for the Bloch-Redfield approach used below, and we will assume that the main effect of $H_2$ is to contribute to the fast quasi-particle relaxation processes.

\begin{figure}
\centering
\includegraphics[width=0.25\textwidth]{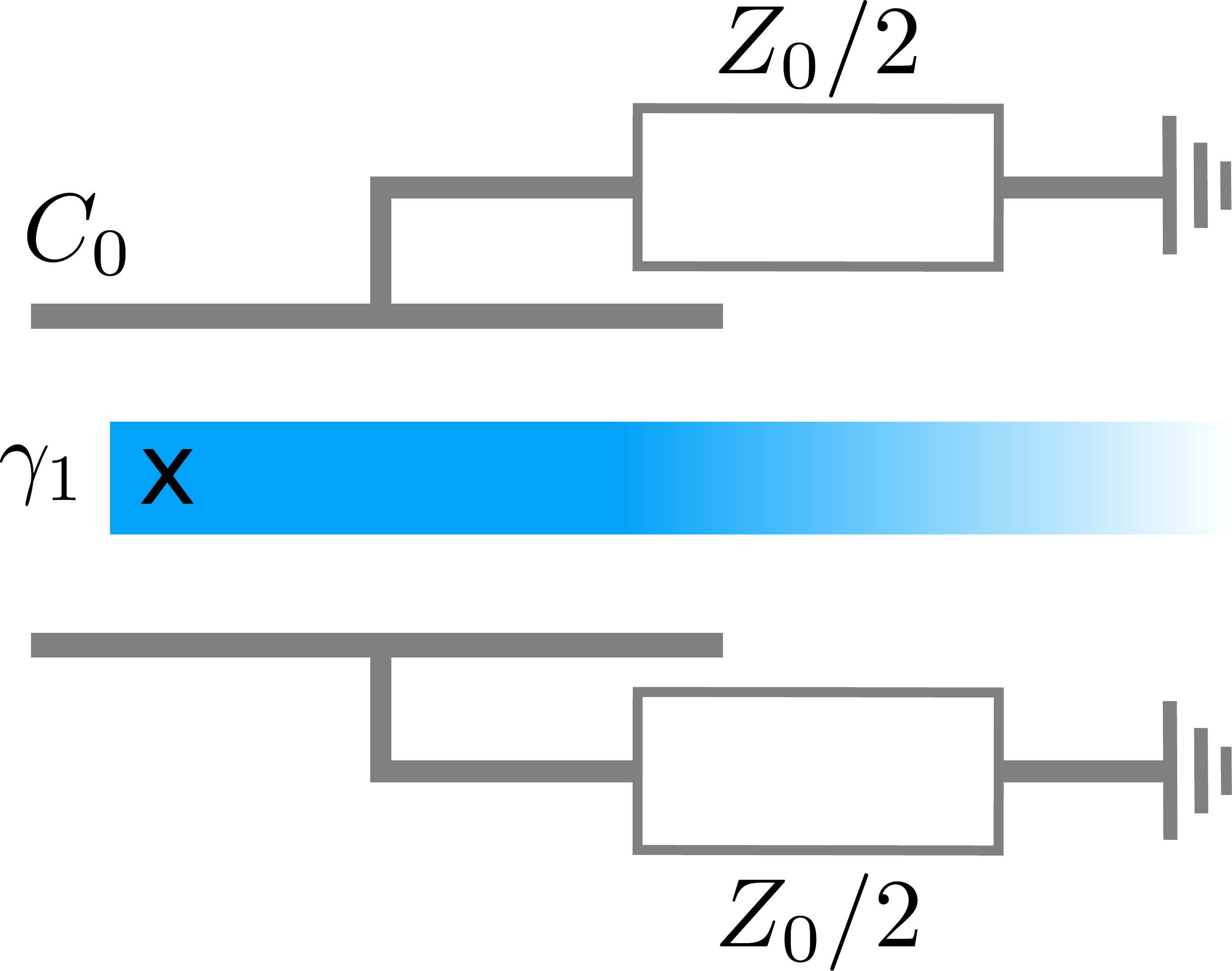}
\caption{Equivalent circuit for the electromagnetic environment coupled to the Majorana operator $\gamma_1$. The environments near the other MBSs are not shown.  \label{fig3}}
\end{figure}

Finally, the Gaussian Hamiltonian $H_\varphi$ is fully determined by first noting that $\langle \varphi_j \rangle^{}_{H_\varphi}=0$ and then specifying the two-point bath correlation function  \cite{Weiss2010}.
For simplicity, we here assume that the different environments in the various regions ($j=1,\ldots,M$) can be characterized by the same spectral density  $J(\omega)$.
By assumption (\textit{iv}) above, the only non-vanishing correlator is given by
\begin{eqnarray}\label{Bidef}
  B(t) &=&  \left\langle \varphi_j(t) \varphi_j(0)\right\rangle^{}_{H_\varphi}  \\ \nonumber
  &=&\int_0^\infty\frac{ d\omega}{2\pi}\, J(\omega) \left[e^{-i\omega t}\left(1+\nB(\omega)\right)+e^{i\omega t}\nB(\omega)\right],
\end{eqnarray}
where $\nB(\omega)=1/(e^{\beta\omega}-1)$ with $\beta=1/\kB T$ is the Bose-Einstein function.
The spectral density $J(\omega)$ of the electromagnetic environment is taken for the equivalent circuit in Fig.~\ref{fig3},
where  fermions couple through the capacitance $C_0$ to the electromagnetic environment with resistance $Z_0$. We note that other spectral densities, for example, containing a $1/f$ component could be more relevant, but here we focus on the so-called Ohmic case for simplicity. Using linear response theory, $B(t)$ in Eq.~\eqref{Bidef} can be related to the impedance of the circuit \cite{Weiss2010}. We thereby obtain the spectral density
\begin{equation} \label{spectralfunction}
    J(\omega) = \frac{2e^2 \omega_0}{C_0}\frac{\omega}{\omega^2+\omega_0^2},
    \quad \omega_0=\frac{1}{C_0 Z_0}.
\end{equation}
{The linear low-frequency dependence is characteristic of an Ohmic environment. It is of course possible to engineer the environment spectral function, such that it is has low-energy modes suppressed. This would be relevant if one wants to improve on the adiabaticity conditions for the qubit operations.}

\section{Readout of the Majorana qubit with environmental coupling}
\label{sec:readout}

\subsection{General remarks}

In this subsection, we discuss on a general level the principles of  Majorana-qubit readout when the qubit Hamiltonian does not commute with the degree of freedom that is being measured. As we saw above, the qubit in principle still has topological protection in the sense that the total-parity qubit operators $\tilde\sigma_i$ in Eq.~\eqref{taupaulidefs} are conserved and cannot be measured by any local operator (when the readout device is detached). However, when attached to the readout device, the topological protection is of course broken and care must be taken if the measurement device should not give the wrong readout yielding a loss of fidelity.

We define a basis $\ket{p,\tilde\sigma}$ using eigenstates of the Pauli operator $\tilde\sigma_z=i\tilde{\gamma}_1\tilde{\gamma}_3$ which is the basis natural for the quantum-dot coupling in Fig.~1,
\begin{equation}\label{totalparityeigenbasis}
  i\tilde{\gamma}_1\tilde{\gamma}_3\ket{p,\tilde\sigma}=\tilde\sigma\ket{p,\tilde\sigma},
\end{equation}
where $p$ refers to environmental quantum numbers (see below for a concrete calculation to first order). The states $\{\ket{p,\tilde\sigma}\}$ are also eigenstates of the Hamiltonian
\begin{equation}\label{Hqubitintotalparityeigenbasis}
  H_\mathrm{qubit}=\sum_{p,\tilde\sigma=\pm1}\Omega_p \ketbra{p,\tilde\sigma}{p,\tilde\sigma},
\end{equation}
where $\Omega_p$ are the eigenenergies.
The even and odd eigenstate sectors are related by
\begin{equation}\label{fromneta1tominusone}
 \ket{p,-1}=\tilde{\gamma}_1\ket{p,1}.
\end{equation}

Next, we wish to express the operator $\sigma_z=i\gamma_1\gamma_3${, which couples to the quantum dot, see Eq.~\eqref{Energygeneralnoqubit},} in the eigenbasis of the topological qubit. First, we note that
\begin{subequations}
\begin{align}\label{gamma12matrixelementa}
   \matelem{p,-1}{i\gamma_1\gamma_3}{p',-1}&=-\matelem{p,1}{i\gamma_1\gamma_3}{p',1},\\
   \label{gamma12matrixelementb}
   \matelem{p,-1}{i\gamma_1\gamma_3}{p',1}&=0.
\end{align}
\end{subequations}
The first relation follows from Eq.~\eqref{fromneta1tominusone} and the definition of $\tilde\gamma_1$ in Eq.~\eqref{tildegammedeff}, while the second one follows from parity conservation. These relations now allow us to write the operator $i\gamma_1\gamma_3$ as
\begin{equation}\label{gamma12inHeigenstates}
{\sigma}_z=i\gamma_1\gamma_3=\sum_{pp'}A_{pp'}\big[\ketbra{p,1}{p',1}-\ketbra{p,-1}{p',-1}\big].
\end{equation}
For an example of $A_{pp'}$, see below where we calculate it in perturbation theory.

Let us now discuss the readout procedure using a quantum dot that effectively couples to the operator in Eq.~\eqref{gamma12inHeigenstates}. Clearly the bare-Majorana Pauli operator ${\sigma}_z$ does in general not commute with the Hamiltonian of the qubit, $ H_\mathrm{qubit}$, in Eq.~\eqref{Hqubitintotalparityeigenbasis}. However, if we consider the situation where the energy scales of the quantum dot, the inverse time scales for switching on the readout circuit, $\tau^{-1}$, and temperature $\kB T$, all are well within the gap of the topological superconductor  $(\varepsilon_d,t_1,t_2,\tau^{-1},\kB T)\ll\Delta$, we should project Eq.~\eqref{gamma12inHeigenstates} to the low-energy sector determined by these energy scales. Moreover, for the case without splitting of the topological qubit, the initial density matrix of the qubit is assumed to be in a thermal (low-temperature) state of the form
\begin{equation}\label{densmatqubitthermal}
  \rho_\mathrm{qubit}=\sum_p\big(\alpha\ket{p,1}+\beta\ket{p,-1}\big)
  \big(\alpha^*\bra{p,1}+\beta^*\bra{p,-1}\big)P_p,
\end{equation}
which has full coherence in the topologically protected sector. Here $P_p\propto \exp(-\Omega_p/\kB T)$ is the thermal distribution.

\begin{figure}
\includegraphics[width=0.4\textwidth]{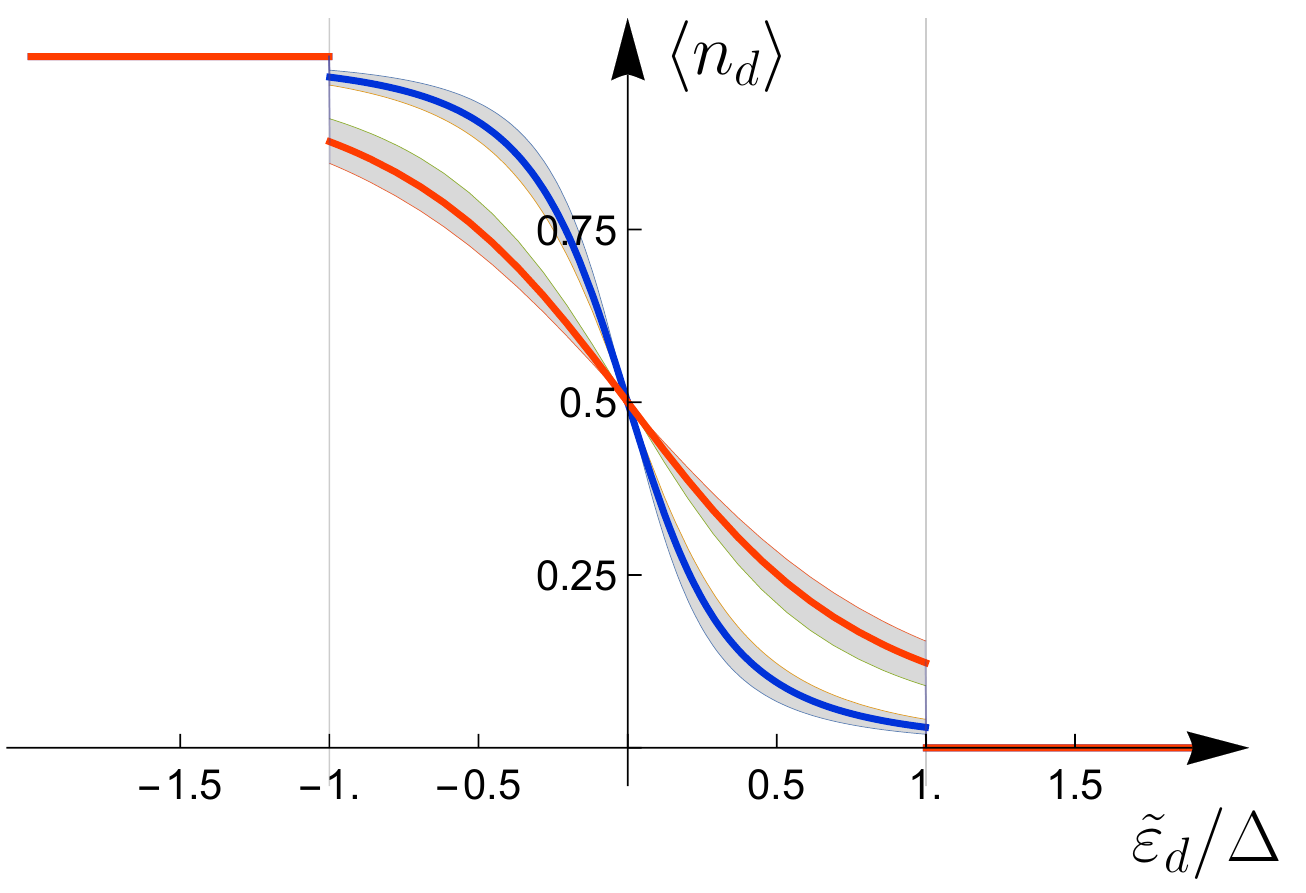}
  \caption{\label{fig:meas} The measured dot occupation $\average{n_d}$ as a function of the dot potential $\tilde\varepsilon_d$ for the readout protocol shown in Fig.~1. The outcomes for the two parity states are illustrated with the blue and red curves. The full lines show the result without environmental coupling, while the shaded areas illustrate the possible outcomes for when the bare-Majorana parity $i\gamma_1\gamma_3$ does not commute with the qubit Hamiltonian. If the shaded regions do not overlap, the coupling to environments gives rise to a visibility reduction, while when they do overlap the fidelity is reduced. Note that $|\tilde\varepsilon_d|$ must be smaller than the gap (vertical lines) for the readout to be valid.}
\end{figure}

When adding the measurement circuit, the Hamiltonian has additional terms. For example, for weak dot tunneling and $\tilde\varepsilon_d>0$, the Hamiltonian is given by Eq.~\eqref{HdotcoupleMBS2ndorder}, which in the qubit eigenbasis follows from Eqs.~\eqref{Hqubitintotalparityeigenbasis} and \eqref{gamma12inHeigenstates},
\begin{equation}\label{Hdotqubitpbasis}
  H^{(2)}=\sum_{pp',\tilde\sigma}\left(\Omega_p\delta_{pp'}+\tilde\sigma A_{pp'} \Lambda\right)\ketbra{p,\tilde\sigma}{p',\tilde\sigma}
  -\frac{|t_1|^2+|t_3|^2}{\tilde{\varepsilon}_d},
\end{equation}
where we defined the energy scale that splits the topological degeneracy
\begin{equation}\label{Lambdadef}
\Lambda = \frac{2\mathrm{Im}[t_1t_3^*]}{\tilde{\varepsilon}_d}.
\end{equation}
The sensor measures the charge on the dot, which is given by the operator $n_d=dH^{(2)}/d\tilde\varepsilon_d$.

After the quantum-dot read-out circuit has been switched on, the population of the energy spectrum is not necessarily a thermal population of the eigenstates as in Eq.~\eqref{Hdotqubitpbasis}, but depends on the protocol for attaching the dot. The readout fidelity then depends on both this population and on the structure of the matrix $\mathbf{A}$.

For small dot-qubit coupling (which means that $\Lambda$ is small compared to $\kB T$ and $\hbar\tau^{-1}$), one can treat the parity dependent term in Eq.~\eqref{Hdotqubitpbasis} as a perturbation. Therefore, if the diagonal elements $A_{pp}$ for the relevant energies have a definite sign, the total-parity degree of freedom $\eta$ can in principle be read out with perfect fidelity. This corresponds to the situation depicted in Fig.~\ref{fig:meas} where the possible outcomes for the two parity states do not overlap. In this situation, the environmental coupling only leads to a reduction of visibility. However, if the sign of  $A_{pp}$ varies for the populated energies, the two distribution functions of possible readouts overlap and, as a result, the fidelity is reduced.

For stronger $\Lambda$, the full matrix $A_{pp'}$ is important for determining the eigensystem of the Hamiltonian \eqref{Hdotqubitpbasis}. Again, if the signs of the diagonal elements of $A$ in this new basis are {not} unique for the energies that are populated, the fidelity of the readout procedure is reduced.

\subsection{Perturbative treatment of the entangled environment-qubit basis}
\label{sec4}

In this subsection, we present a perturbative analysis of the entanglement between the MBS manifold and the environment and discuss the consequences for the readout visibility/fidelity. For concreteness, we discuss the case $M=4$ and use the basis $|n_{13},n_{24}\rangle\otimes|\{k,j\},\{q,j\}\rangle_\mathrm{env}$, where $\{k,j\},\{q,j\}$ label states in the quasi-particle and boson environments, respectively. On the other hand,  ${n_{13}}$ and  ${n_{24}}$ refer to the fermion level occupations
corresponding to the respective fermion operators,
\begin{equation}\label{auxferm}
{d_{13}=(\gamma_1+i\gamma_3)/2, \quad d_{24}=(\gamma_2+i\gamma_4)/2,}
\end{equation}
where the number states $|n_{13}, n_{24}\rangle$ follow from the empty state, $|00\rangle$, as
\begin{equation}\label{basis1}
d_{13}^\dag|00\rangle=|10\rangle, \quad d_{24}^\dag|00\rangle=|01\rangle, \quad d_{24}^\dag d_{13}^\dag|00\rangle=|11\rangle.
\end{equation}
The bosonic environment is in  diagonal form written as
\begin{equation}\label{Hphidef}
  H_\varphi=\sum_{q,j}\omega_{q,j}^{\phantom{\dagger}} b_{q,j}^\dag b_{q,j}^\nodag,
\end{equation}
where $b_{q,j}$ are boson annihilation operators. The potential fields $\varphi_j$ are given in terms of the bosons as
\begin{equation}\label{varphidef}
  \varphi_j=\sum_{q}\left(M_{q,j}^* b_{q,j}^\nodag+M_{q,j}^\nodag b_{q,j}^\dag\right).
\end{equation}
For the arguments in this subsection, we do not need the explicit form of the matrix elements $M_{q,j}$.

We take the case of even total fermion-number parity. The two degenerate even-parity unperturbed ground states are $|00\rangle\otimes |0\rangle_\mathrm{env}$ and $|11\rangle\otimes |0\rangle_\mathrm{env}$. First order perturbation theory then gives that the ground states of the interacting system are
\begin{equation}\label{newGS}
\begin{aligned}
\ket{G_0} = &\frac1{\sqrt{C}} \Big(\ket{00}\ket{0}_\mathrm{env}-\sum_{ikq}\frac{W_{k,i}M_{q,i}}{E_k+\omega_q}\gamma_i\ket{00}\alpha^\dag_{k,i}b^\dag_{q,i} \ket{0}_\mathrm{env}\Big),\\
\ket{G_1} =& \frac1{\sqrt{C}} \Big(\ket{11}\ket{0}_\mathrm{env}-\sum_{ikq}\frac{W_{k,i}M_{q,i}}{E_k+\omega_q}\gamma_i\ket{11} \alpha^\dag_{k,i}b^\dag_{q,i}\ket{0}_\mathrm{env}\Big),
\end{aligned}
\end{equation}
with
\begin{equation}\label{Adef}
  C=1+\sum_{ikq}\frac{|W_{k,i}M_{q,i}|^2}{(E_k+\omega_q)^2}\equiv 1+\eta.
\end{equation}
For simplicity, we here assume that the energies $E_k$ and $\omega_q$ are  identical in different regions, $j=1,\ldots,4$.

The excited states can be written down in a similar way. For example, let us consider the unperturbed excited states $b^\dag_{q,j}\ket{ss}\otimes|0\rangle_\mathrm{env}$ (with $s=0,1$), where the corresponding entangled excited states are to first order given by
\begin{equation}\label{newES}
\begin{aligned}
\ket{E_{qjs}} = & \frac1{\sqrt{B_{q,j}}}\Big(b_{q,j}^\dag\ket{ss}\ket{0}_\mathrm{env}\\
&-\sum_{ikq'}\frac{W_{k,i}M_{q',i}}{E_k+\omega_{q'}}\gamma_i\ket{ss}\alpha^\dag_{k,i}b^\dag_{q'i}b_{q,j}^\dag\ket{0}_\mathrm{env}\\
&-\sum_{k}\frac{W_{k,j}M^\ast_{q,j}}{E_k-\omega_{q}}\gamma_j\ket{ss}\alpha^\dag_{k,j}\ket{0}_\mathrm{env}\Big),
\end{aligned}
\end{equation}
with the normalization factor
\begin{equation}\label{Bdef}
  B_{q,j} = 1 + \sum_{ikq'}\frac{|W_{k,i}M_{q',i}|^2}{(E_k+\omega_{q'})^2}
  +\sum_{k,z=\pm1}\frac{|W_{k,j}M_{q,j}|^2}{(E_k+z\omega_{q})^2}.
\end{equation}
Similarly, one can generate the corrections to the unperturbed two-boson excited states: $b^\dag_{q,j}b^\dag_{q',j'}\ket{ss}\otimes|0\rangle_\mathrm{env}$, etc.

With the above perturbative results for the eigenstates, one can now construct the corresponding matrix elements of the matrix $\mathbf{A}$. As we saw in Eq.~\eqref{gamma12inHeigenstates}, the matrix elements are identical for the two topologically degenerate sectors. For example, we have for the diagonal elements
\begin{eqnarray}
  A_{G_0,G_0} &=& A_{G_1,G_1}= \expect{G_0}{i\gamma_1\gamma_3},\\
  A_{q10,q10} &=& A_{q11,q11}=\expect{E_{q10}}{i\gamma_1\gamma_3},
\end{eqnarray}
that
\begin{equation}
  A_{G_0,G_0} =\frac{2-C}{1+C}=\frac{1-\eta}{2+\eta}, \quad  A_{q10,q10} =\frac{2-B_{q,1}}{1+B_{q,1}}.
\end{equation}
Off-diagonal elements between the excited states in Eq.~\eqref{newES} are
\begin{equation}\label{offdiagonalA}
  A_{q10,q'10}=-\sum_{k,z=\pm1}\frac{|W_{k,1}^\nodag|^2M^\nodag_{q,1}M^*_{q',1}}{(E_k+z\omega_q)(E_k+z\omega_{q'})}
\end{equation}
for $q\neq q'$.

From the above, we conclude that for weak measurements and weak coupling to the environment, there is no loss of fidelity (only visibility) as long as the diagonal elements remain positive, whereas for strong measurement one has to investigate the structure of the eigenvalue spectrum of $\mathbf{A}$ more carefully. However, it is not well understood when the assumption of weak measurements is valid. It depends on both the strength of $\Lambda$ and the time scale $\tau$, and it is an interesting topic for further studies. Here we focus on establishing the result for the limit of an instantaneous and projective measurement of the bare MBSs, \textit{i.e.}, the bare Pauli operator $\sigma$. This is the topic of the next section.

\section{Readout of a Majorana basis in the sudden approximation}
\label{sec:readoutsudden}

In this section, we discuss the limit where the operator $\sigma_z=i\gamma_1\gamma_2$ is measured projectively. This is relevant when the energy scale $\Lambda$ in Eq.~\eqref{Hdotqubitpbasis} is larger than temperature and the time scale $\tau$ for turning on the measurement device is short compared to all scales, including $\Delta$. As discussed above this therefore constitutes a worst case scenario and slower turn-on would reduce the fidelity loss even though full adiabaticity is never possible because the combined fermion-boson system is gapless.

\subsection{Perturbative estimate}

Let us start by assuming that the system has been initialized in a linear superposition of the two dressed ground states \eqref{newGS} (at $T=0$),
\begin{equation}\label{GSsup}
\ket{\psi}=\alpha\ket{G_0}+\beta\ket{G_1}.
\end{equation}
A projective measurement of $\sigma_{z}$   then yields the outcome $+1$  with probability
\begin{equation}\label{project}
  P(\sigma_{z}=1) =\mathrm{Tr}\left(\Pi_1\ket{\psi}\bra{\psi} \right),
\end{equation}
where the projection operator $\Pi_1$ is
\begin{equation}\label{Pidef}
   {\Pi_1 =\sum_{n_{24}=0,1} \ket{0,n_{24}}\bra{0,n_{24}}\otimes \mathds{1}_\mathrm{env},}
\end{equation}
and $\mathds{1}_\mathrm{env}$ denotes the identity operator in the Hilbert space of the environment. The probability in Eq.~\eqref{project} thus becomes
\begin{equation}\label{project2}
  P(\sigma_{z}=1) = \frac{|\alpha|^2}{1+\eta} + \frac{|\beta|^2\eta}{1+\eta},
\end{equation}
and similarly for the probability to measure $\sigma_{z}=-1$,
\begin{equation}\label{project2m1}
  P(\sigma_{z}=-1) = \frac{|\beta|^2}{1+\eta} + \frac{|\alpha|^2\eta}{1+\eta},
\end{equation}
where the decoherence parameter $\eta$ has been defined in Eq.~\eqref{Adef}.
Equations~\eqref{project2} and \eqref{project2m1} show that the readout error is of order $\eta$. Moreover,
because there is no value of $\alpha$  for which $P(\sigma_{z}=1)=1$,
they also demonstrate that reading out $\sigma_{z}$ does not simply correspond to reading out the qubit defined by the basis states $\{\ket{G_0},\ket{G_1}\}$ in some other direction.

At finite temperature, we have instead of Eq.~\eqref{GSsup} a mixed state with contributions from excited states as in Eq.~\eqref{newES}. The resulting density matrix is still coherent within the topologically protected set of degenerate states because no local perturbation (say, for region $j=1$)  mixes the two sectors  $\{\ket{00,\mathrm{even}},\ket{10,\mathrm{odd}}\}$ and $\{\ket{11,\mathrm{even}},\ket{01,\mathrm{odd}}\}$, where odd and even refer to the parity of the quasi-particle continua. However, even though  coherence in the topologically protected subspace is maintained,  the coefficients $\alpha$ and $\beta$ can again not be read out truthfully using the projection \eqref{Pidef} because the projection operators $\Pi_{\pm 1}$ do not commute with the interacting Hamiltonian.

To summarize this subsection, the reduction factors in Eqs.~\eqref{project2} and \eqref{project2m1} are caused by reading out in the bare (undressed) basis $\{\ket{00},\ket{11}\}$ instead of using the true (dressed) states \eqref{newGS}.  The factor $1/(1+\eta)$, which here was determined by first-order perturbation theory, will appear in the non-Markovian Bloch-Redfield approach below, see Sec.~\ref{sec5c}. We note that a similar dressing of the ground state by  environmental modes has been studied in detail for the related but simpler spin-boson model \cite{Weiss2010,LeHur2008}, where the coherence reduction is well established even at zero temperature.

\subsection{Bloch-Redfield approach to sudden readout of a bare-Majorana qubit}
\label{sec5}

We now study the decoherence dynamics of Majorana qubits in terms of a modified Bloch-Redfield approach. The main difference between our approach and standard quantum master equations for, e.g., a qubit coupled to a bosonic bath \cite{Weiss2010,BreuerPetruccione}, arises from the fact that the fermion numbers in the Majorana sector and in the environment are not independent since the total fermion number parity of each spatial region ($j=1,\ldots,M$) is conserved by the full Hamiltonian $H$. In this subsection, we discuss the Bloch-Redfield approach for the general class of models in Sec.~\ref{sec3}. In Sec.~\ref{sec6}, we will then apply these results to a specific TS wire model.

By adopting the standard derivation of quantum master equations \cite{BreuerPetruccione} to the case of our Hamiltonian $H$, we obtain the equation of motion for the reduced density matrix, $\rrho_M(t)$, describing the MBS sector,
\begin{equation}\label{quantummaster}
  \frac{d}{dt}\rrho_M(t) = -\int_0^tdt' \, \Tr\left[H_{\rm int}(t),[H_{\rm int}(t'),\rrho(t')]\right].
\end{equation}
For $M=4$, the space spanned by the MBSs is equivalent to two fermions and  $\rrho_M$ can be represented by a $4\times 4$ matrix.
In Eq.~\eqref{quantummaster},
$H_{\rm int}(t)$ is the MBS-environment coupling Hamiltonian in the interaction picture, with $H_0+H_\varphi$ as  unperturbed part,
 $\rrho(t)$ is the full density matrix of the entire system, and Tr$_{\rm env}$ indicates a trace over environmental degrees of freedom.
In  Eq.~\eqref{quantummaster},
we assume the weak MBS-environment coupling limit such that the standard Born approximation applies \cite{Weiss2010,BreuerPetruccione}.

If relaxation processes in the environment are much faster than the time scale for changes in the reduced density matrix $\rrho_M$,
the density matrix $\rrho(t')$ appearing in Eq.~\eqref{quantummaster} effectively separates into $\rrho_M(t')$ and an environmental part, and we can  neglect MBS-environment entanglement in $\rrho(t')$. Assuming that above-gap quasi-particles quickly decohere because of $H_2$ in Eq.~\eqref{H2def},
$\rrho(t')$ will therefore factorize into  $\rrho_M(t')$ and an \emph{equilibrium} environmental density matrix $\rrho_{\rm env}$.
Since the main role of $H_2$ is to decohere quasi-particles, we also replace $H_{\rm int}(t)\to H_1(t)$ [see Eq.~\eqref{H1def}] in Eq.~\eqref{quantummaster}.

However, there is an important  catch:  the parities of the Majorana subsystem and of the environmental sector are not independent because of total  parity conservation.
 In what follows, we always take the conserved fermion number parity of the entire system as even
such that
\begin{equation}\label{rhosseparate}
  \rrho(t')=\rrho_M^e(t')\otimes \rrho_\env^e+\rrho_M^o(t')\otimes \rrho_\env^o,
\end{equation}
where the superscripts $e/o$ refer to even/odd parity sectors of the respective subsystem.
Next we insert Eq.~\eqref{rhosseparate} into  Eq.~\eqref{quantummaster}.
Noting that coherent contributions with different parities in the Majorana sector are absent, we obtain
\begin{widetext}
\begin{equation}\label{rhoMdiffs}
\frac{d}{dt}\rrho_M^{e/o}(t)=-\sum_{i,j} \int_0^tdt' \left[ g_{ij}^{e/o}(t-t')\gamma_i\gamma_j \rrho_M^{e/o}(t')+g_{ij}^{e/o}(t'-t) \rrho_M^{e/o}(t')\gamma_i\gamma_j-\left(g_{ij}^{o/e}(t-t')+g_{ij}^{o/e}(t'-t)\right)\gamma_i\rrho_M^{o/e}(t^\prime)\gamma_j\right]
\end{equation}
\end{widetext}
with the functions ($i,j=1,\ldots,M)$
\begin{equation}\label{gijdef}
g_{ij}^{e/o}(t-t')=-\left\langle \Gamma_i(t)\varphi_i(t) \Gamma_j(t')\varphi_j(t')\right\rangle_{e/o},
\end{equation}
where $\langle\cdots\rangle_{e/o}={\rm Tr}_{\rm env}\left(\rrho_{\rm env}^{e/o} \cdots\right)$ and $\Gamma_i(t)$ has been defined in Eq.~\eqref{H1def}.

We now use two properties of the environment which follow from the conditions specified after Eq.~\eqref{H2def}. First, all MBSs are assumed to be so far away from each other that there is no phase coherence between quasi-particles in different regions. As a consequence, $g_{ij}\propto \delta_{ij}$. (Nonetheless, quasi-particles may incoherently diffuse throughout the device.)   Second, quasi-particles and bosonic modes are taken to be uncorrelated, implying that the expectation value \eqref{gijdef} can be factorized. This assumption is equivalent to disregarding the Hamiltonian $H_2$  when evaluating $g_{ij}^{e/o}(t)$. (As   discussed above, the main role of $H_2$ is to induce quasi-particle relaxation.)  After those steps, we obtain
\begin{equation}\label{Fijdef}
g_{ij}^{e/o}(t-t')=F_{i}^{e/o}(t-t')B(t-t') \delta_{ij},
\end{equation}
with the boson correlation function $B(t)$ in Eq.~\eqref{Bidef} and the quasi-particle correlator
\begin{eqnarray}\nonumber
 F_{i}^{e/o}(t) &=& -\left\langle \Gamma_i(t)\Gamma_i(0)\right\rangle_{e/o}
  =\int_\Delta^\infty dE\, \nu(E) |W_i(E)|^2
  \\  \label{Fidef} &\times&
  \left[e^{-iEt}(1-\nF^{e/o}(E))+e^{iEt}\nF^{e/o}(E)\right].
\end{eqnarray}
Here  $\nu(E)=\sum_k \delta(E-E_k)$ is the quasi-particle density of states.   From Eq.~\eqref{Wkirdef}, we then obtain
\begin{equation}\label{widef}
  \nu(E) |W_i(E)|^2= \sum_k \delta(E-E_k)  |W_{k,i}|^2.
\end{equation}
The Fermi-Dirac functions in Eq.~\eqref{Fidef} are given by
\begin{equation}\label{nFeo}
  \nF^{e/o}(E)=\frac{1}{e^{\beta (E\pm\delta F)}+1},
\end{equation}
where $\delta F$ is the free-energy difference between the even and odd parity cases, $\delta F=F_\mathrm{odd}-F_\mathrm{even}$.
The thermodynamics of a superconducting island with fixed total parity has been considered in Refs.~\cite{Lafarge1993,Tuominen1993,Higginbotham2015}.
At low temperatures, one can parameterize $\delta F$ by the number $N_{\rm eff}$ of quasi-particle states on the island,
\begin{equation}\label{Neffdif}
 \delta F=\Delta-\kB T \ln N_\mathrm{eff},\quad N_\mathrm{eff}\simeq \int_\Delta^\infty dE\, \nu(E) e^{-\beta(E-\Delta)}.
\end{equation}
Assuming a BCS form for $\nu(E)$, one obtains the estimate
\begin{equation} \label{Neff2}
    N_\mathrm{eff}\approx d_\mathrm{S}V_\mathrm{S} \sqrt{2\pi \kB T\Delta},
\end{equation}
where $d_\mathrm{S}$ is the normal density of states and $V_\mathrm{S}$ the volume of the superconductor. We note that $N_\mathrm{eff}$ determines the temperature $T^*$ at which the probability of having the first quasi-particle in the system approaches unity, $T^*\approx \Delta/(\kB N_\mathrm{eff})$. Recent experiments have reported the value $T^*\approx$ 0.3K for a single nanowire \cite{Higginbotham2015}.

\subsection{Markovian approximation}
\label{sec5b}

The integro-differential equation \eqref{rhoMdiffs} includes memory effects because the change of  $\rrho_M(t)$ depends on  $\rrho_M(t')$ at earlier times, $t'<t$. One can in principle solve this equation but in order to have simple results (and to reproduce results obtained by earlier studies), we first turn to the  Markovian  approximation.
The standard Markovian approximation for the Bloch-Redfield master equation \eqref{rhoMdiffs} involves two steps \cite{Weiss2010,BreuerPetruccione}. First, the density matrix $\rrho_M(t')$ under the integral is replaced by $\rrho_M(t)$.  Second, the upper limit in the time integral is replaced by infinity.
In addition, to simplify notation, we again take identical but uncorrelated environments for different MBSs. With these steps,  the master equation  \eqref{rhoMdiffs} is given in Lindblad form,
\begin{equation}\label{rhodiffmarkov}
  \frac{d}{dt}\rrho_M^{e/o}=  -\Gamma^{e/o}\rrho_M^{e/o}+\frac{\Gamma^{o/e}}{M}\sum_i\gamma_i^\notop\rrho_M^{o/e}\gamma_i^\notop,
\end{equation}
with the rates [cf.~Eq.~\eqref{Fijdef}]
\begin{equation}\label{disscons}
\Gamma^{e/o}=M\int_{-\infty}^{\infty}dt\,g^{e/o}(t) = M
\int_\Delta^\infty dE f^{e/o}(E).
\end{equation}
We here define the auxiliary functions
\begin{equation}\label{feo}
f^{e/o}(E) = \nu(E) \abs{W(E)}^2 J(E) \left( \nB(E) + \nF^{e/o}(E)\right).
\end{equation}
For low temperatures, $T\ll T^*$, we now have
\begin{equation}
\nB(E)+\nF^{e/o}(E)\simeq
\left\{ \begin{array}{cc} e^{-\beta E}
,& {\rm even},\\
N_{\rm eff}^{-1} e^{-\beta (E-\Delta)}
,& {\rm odd}.
\end{array}\right.
\end{equation}
From Eq.~\eqref{disscons}, we thus obtain the asymptotic low-temperature expressions
\begin{eqnarray} \label{lowtemprates}
  \Gamma^{o}&\approx&  \kB T N^{-1}_{\rm eff} {\cal S}(\Delta) ,
  \\ \nonumber
  \Gamma^{e} &\approx& \kB T {\cal S}(\Delta) e^{-\Delta/\kB T},
\end{eqnarray}
with ${\cal S}(\Delta)=M \nu(\Delta) |W(\Delta)|^2 J(\Delta)$.
We observe that in general, $\Gamma^o\gg\Gamma^e$ due to the absence of the exponential suppression factor in $\Gamma^o$.
To understand this result, note that for even total parity, the odd parity Majorana sector must come with at least one quasi-particle excitation.
For $T>0$, this above-gap excitation can now quickly relax and thereby bring the Majorana subsystem to the energetically favorable even parity sector.

To explicitly obtain the decoherence dynamics from the Lindblad equation   \eqref{rhodiffmarkov}, we take  $M=4$ and  parametrize  $\rrho_M^{e/o}$  in the basis introduced in Eqs.~\eqref{auxferm} and \eqref{basis1}. With real coefficients {$a_\pm^{e/o}$} and complex numbers $b^{e/o}$,
{
\begin{equation}\label{rhoevenoddparam}
  \rrho_M^{e/o}=  \left(  \begin{array}{cc} a_+^{e/o} &  b^{e/o}\\     (b^{e/o})^* & a_-^{e/o}     \end{array} \right),
\end{equation}
where $p^{e/o}=a_+^{e/o}+a^{e/o}_-$} is the probability for the Majorana sector having even/odd parity, respectively.
We next note that
the last term in Eq.~\eqref{rhodiffmarkov} can be written as
\begin{equation}\label{gammarhogamma}
  \sum_i\gamma_i^\notop\rrho_M^{e/o}\gamma_i^\notop= 2p^{e/o} P_{o/e},
\end{equation}
where  $P_{o/e}$ is the projector onto the odd/even parity Majorana subspace. {The identity \eqref{gammarhogamma} follows directly by using the basis defined in Eq.~\eqref{basis1} along with the definition of $d_{13}$ and $d_{24}$ in Eq.~\eqref{auxferm}.} We will see below that  for $t\to \infty$ and $T>0$,
Eq.~\eqref{gammarhogamma} implies that the bare-Majorana qubit will fully decohere.
The simple form of Eq.~\eqref{gammarhogamma} is a consequence of our assumption that different environments are identical and uncorrelated. If they have different spectral functions, the long-time limit of $\rrho_M(t)$ is also affected.

Let us now assume that at time $t=0$, we start from the even parity Majorana sector,
i.e., $\rrho_M^o(0)=0$.  The
off-diagonal components of  $\rrho_M^{e}$ will then show an exponential decay with rate $\Gamma^e$,
\begin{equation}\label{rhooffdiagonal}
  b^e(t)=  e^{-\Gamma^e t} \,b^e(0).
\end{equation}
{Using the normalization condition $p^e + p^o = 1$, the} dynamics of the diagonal elements
{$a_\pm^e$} follows from {
\begin{equation}\label{diffad}
  \dot{a}_\pm^e=-\Gamma^e a_\pm^e + \frac{\Gamma^o}{2} (1-a_+^e-a_-^e).
\end{equation}  }
By adding those equations, we obtain
\begin{equation}\label{diffevenprob}
  \dot{p}^e= -\Gamma^e{p}^e+ \Gamma^{o} (1-{p}^e),
\end{equation}
with the solution
\begin{equation}\label{evenprobsol}
  {p}^e(t) = e^{-(\Gamma^e +\Gamma^o)t}(1-p_\mathrm{eq})+p_\mathrm{eq},
\end{equation}
where the equilibrium probability reached for $t\to \infty$ is
 \begin{equation}\label{peq}
   p_\mathrm{eq}=\frac{1}{1+\Gamma^e/\Gamma^o}.
\end{equation}
Inserting Eq.~\eqref{evenprobsol} back into Eqs.~\eqref{diffad} one easily finds {$a^e_+(t)$ and $a_-^e(t)$}, given their initial values at $t=0$. Equation \eqref{evenprobsol} shows that the decay towards equilibrium  involves two separate contributions. One is due to the rate $\Gamma^e$ which is exponentially small at low temperatures. The other is due to $\Gamma^o$  which does not contain the exponential suppression factor and thus implies a faster decay (for $T>0$). In addition, we observe from Eq.~\eqref{peq} that for $\kB T\ll \Delta$, the probability for remaining in the even parity sector at $t\to \infty$ is very close to unity, $p_{\rm eq}\simeq 1-N_{\rm eff}e^{-\Delta/\kB T}$, see Eq.~\eqref{lowtemprates}. In particular, at $T=0$ the bare-Majorana qubit does not decohere at all within the Markovian approximation. This conclusion and some of the above results have been reported before, see, e.g., Refs.~\cite{Goldstein2011,Schmidt2012}.

\subsection{Non-Markovian case}
\label{sec5c}

\subsubsection{The $T=0$ case}

We next turn to the $T=0$ qubit dynamics and take into account non-Markovian memory effects. In Sec.~\ref{sec4}, we have presented a fidelity reduction mechanism for the bare-Majorana qubit state due to entanglement of the MBS sector with environmental degrees of freedom. Within the Markovian approximation, this effect is exponentially suppressed at low temperatures due to the energy difference  $\Delta$ between both sectors. For our system, this conclusion equivalently follows under a Fermi Golden Rule approach with on-shell scattering between the two parity sectors. However, we will show below that the fidelity of the bare-Majorana qubit is affected even at $T=0$ due to \emph{virtual off-shell} processes which give rise to non-Markovian dynamics.

Our starting point is Eq.~\eqref{rhoMdiffs}, where we again assume that the  environments coupled to different MBSs are identical but uncorrelated. Setting $M=4$, we parameterize $\rrho_M^{e/o}$ using  the real Bloch vector components  $d_\alpha^{\,e/o}$
and population factors $p^{e/o}$,
\begin{equation}\label{rhoddef}
\rrho_M^{e/o}(t)=\sum_{\alpha=x,y,z}d_\alpha^{\,e/o}(t) \sigma_\alpha^{e/o} + \frac12 p^{\,e/o}(t) P_{e/o},
\end{equation}
where the Pauli matrices $\sigma^{e/o}_{\alpha}$ act in the even/odd $2\times 2$ spaces defined in Eq.~\eqref{rhoevenoddparam} and
 $P_{e/o}$ projects to the even/odd parity Majorana sector.
From Eq.~\eqref{rhoMdiffs}, we then obtain  the non-Markovian $T=0$ master equation
\begin{equation}\label{rhoMdiffs0}
\frac{d}{dt}\rrho_M^{e/o}(t)=- 4\int_0^tdt'\, g(t-t') \left(\rrho_M^{e/o}(t') -\frac{p^{e/o}(t')}{2}P_{o/e}\right).
\end{equation}
 The function $g(t)=g^{e/o}(t)+g^{e/o}(-t)$ follows from Eq.~\eqref{Fijdef}, where we notice that $g^{e/o}(t)$  does not depend on parity ($e/o$) for $T=0$,
\begin{equation}\label{gzeroTdef}
  g(t)=\frac{1}{\pi}\int_{0}^{\infty}d\omega   \int_\Delta^\infty dE \, \nu(E) |W(E)|^2
   J(\omega)
  \cos[(\omega+E) t].
\end{equation}

Let us first consider  the dynamics of $d_\alpha(t)$. The equations of motion are obtained by multiplying Eq.~\eqref{rhoMdiffs0} with $\sigma_\alpha^{e/o}$ and  taking the trace,
\begin{equation}\label{rhoMdiffspauli}
  \dot d^{\,e/o}_\alpha(t)=- 4\int_0^tdt'\, g(t-t') d_\alpha^{\,e/o}(t').
\end{equation}
The solution follows by Laplace transformation,
\begin{equation}\label{disollaplace}
  \tilde d_\alpha^{\,e/o}(s)= \frac{1}{s+4\tilde g(s)}
  \ d_\alpha^{\,e/o}(t=0),
\end{equation}
where $\tilde h(s)$ denotes the Laplace transform of a function $h(t)$. For the asymptotic long-time behavior, we thereby find
\begin{equation}\label{disollaplacelongtimes}
  d_\alpha^{\,e/o}(t\to\infty)= \frac{1}{1+\eta}
\  d_\alpha^{\,e/o}(t=0),
\end{equation}
with the dimensionless decoherence parameter
\begin{equation}\label{alphadef}
  \eta= \frac{4}{\pi}\int_0^\infty d\omega
  \int_\Delta^\infty dE \frac{J(\omega)\nu(E)|W(E)|^2}{(\omega+E)^2}.
\end{equation}
The coherences encoded by $d_\alpha^{\,e/o}(t)$ are thus reduced for $t\to \infty$ due to the coupling of MBSs to quantum fluctuations of the environment, even at zero temperature.
Quantitatively, this effect is described by the number $\eta$ as explained in Sec.~\ref{sec4}.
Although $d_\alpha^{\,e/o}(t)$ does not decay all the way down to zero for $t\to \infty$, it is reduced by a finite amount.  Note that
this result equally applies  to both  parity sectors.

Likewise, the equations of motion for the population factors follow as
\begin{equation}
  \dot p^{\,e/o}(t)=- 4\int_0^tdt'\, g(t-t') \left[p^{\,e/o}(t')-p^{\,o/e}(t')\right].
\end{equation}
After Laplace transformation, we have
\begin{equation}\label{rhoMdiffstrace}
  s\tilde p^{\,e/o}(s)-p^{\,e/o}(t=0)=- 4\tilde g(s) [\tilde p^{\,e/o}(s)-\tilde p^{\,o/e}(s)].
\end{equation}
Noting that $\tilde p^{\,e/o}(s)+\tilde p^{\,o/e}(s)=1/s$ because of $p^{\,e/o}(t)+p^{\,o/e}(t)=1$, Eq.~\eqref{rhoMdiffstrace} yields
\begin{equation}\label{rhoMdiffstracesol}
  \tilde p^{\,e/o}(s)=\frac{p^{\,e/o}(t=0)+4\tilde g(s)/s}{s+8\tilde g(s)}.
\end{equation}
From this expression, the asymptotic long-time behavior follows in the form
\begin{equation}\label{rhoMdiffstracesollongtimes}
 p^{\,e/o}(t\to\infty)=\frac{p^{e/o}(t=0)+\eta}{1+2\eta}.
\end{equation}
Starting, say, from the even parity sector,  the probability to end up with odd parity is given by $p^{\,o}( \infty)= \eta/(1+2\eta)\le 1/2$.
For $\eta\to \infty$, the full parity mixing limit with $p^{\,e}(\infty)=p^{\,o}(\infty)=1/2$ is realized.  In that case,  also all coherences die out, $d_\alpha^{e/o}(\infty)\to 0$.
Importantly,
these predictions are in marked contrast to the corresponding $T=0$ Markovian results in Sec.~\ref{sec5b}.

\begin{figure}
\centering
\includegraphics[width=0.45\textwidth]{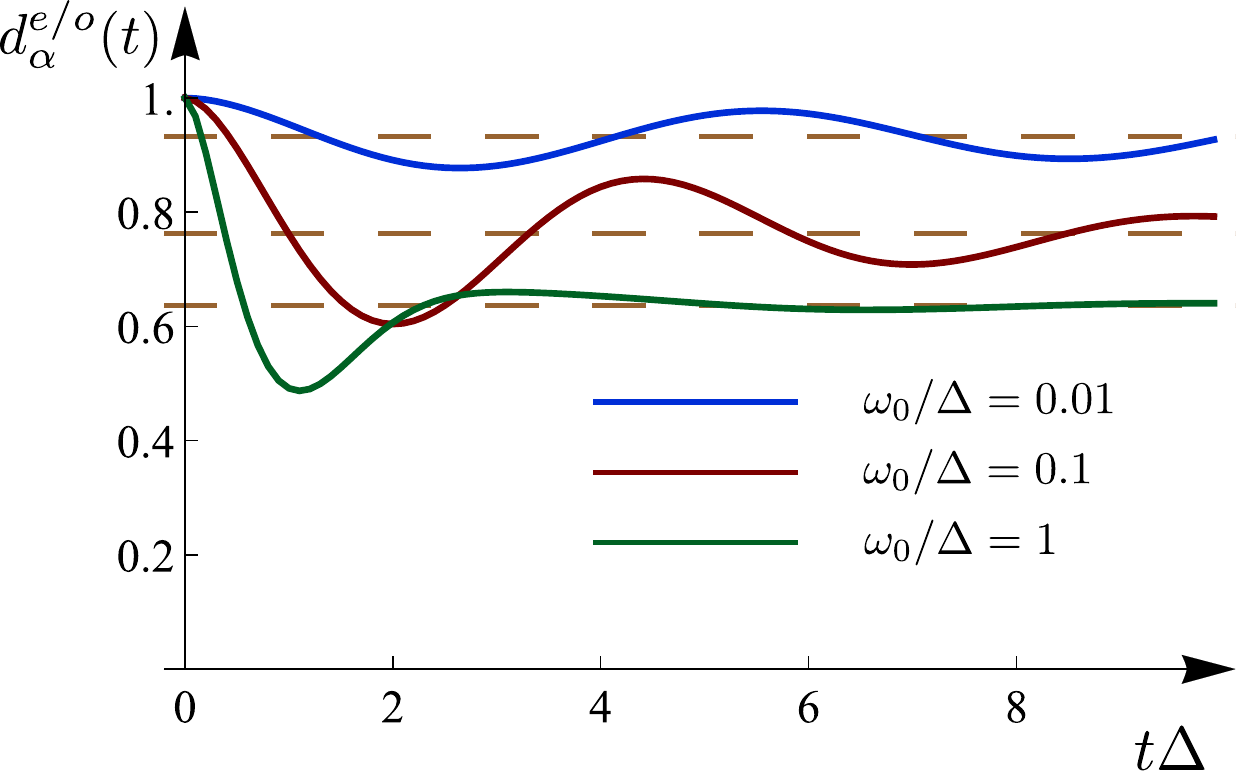}
\caption{Time dependence of the coherences, $d_\alpha^{\,e/o}(t)/d_\alpha^{\,e/o}(0)$, at $T=0$, where results are independent of the parity ($e/o$) sector and of the component ($\alpha=x,y,z$). For three values of $\omega_0/\Delta$, cf.~Eq.~\eqref{spectralfunction}, the curves have been obtained numerically by inverse Laplace transformation of Eq.~\eqref{disollaplace}, with $B=1$ in Eq.~\eqref{disollaplacehs}. Dashed lines show the respective long-time asymptotic value $1/(1+\eta)$. }\label{fig4}
\end{figure}

In order to obtain the full time dependence of the $T=0$ coherences in the non-Markovian case, the inverse Laplace transformation of Eq.~\eqref{disollaplace} has been performed numerically by using a simplifying assumption for the $E$-integral in Eq.~\eqref{gzeroTdef},  replacing $E\to \Delta$ in the cosine. The rationale behind this approximation is that for the
$p$-wave nanowire model in Sec.~\ref{sec6}, the function $\nu(E)|W(E)|^2$ has a clear peak at $E$ slightly above $\Delta$, see Fig.~\ref{fig10} in  App.~\ref{App::WandJ}. Using Eq.~\eqref{spectralfunction}, the $\omega$-integral can then be performed, see App.~\ref{app:laplace} for details.  The corresponding numerical results are shown in Fig.~\ref{fig4} and illustrate how Eq.~\eqref{disollaplacelongtimes} is approached at long times. We observe that the coherences oscillate and decay on time scales corresponding to fractions of $\Delta^{-1}$.
For smaller $\eta$, we find that both the oscillations and the decay become slower.

We  conclude that at $T=0$, non-Markovian effects can be very important. In particular, they induce  a coherence reduction and cause parity mixing between the bare-Majorana qubit and the environment, especially for large $\eta$ in Eq.~\eqref{alphadef}.

\subsubsection{Finite $T$}
\label{sec5d}

For finite $T$, we have to distinguish $g^{e}(t)$ and $g^{o}(t)$.
In the Laplace domain, this parity-dependent correlation function can be calculated for $\text{Re}(s)>0$ and subsequently be analytically continued to $\text{Re}(s)<0$.
From Eq.~\eqref{Fijdef}, we then find
\begin{align}\label{gevenodd}
\tilde g^{\,e/o}(s)
&= \frac{s}{\pi} \int_0^\infty d\omega \int_\Delta^\infty dE \ \nu(E) \abs{W(E)}^2 J(\omega)\\ \nonumber
&\times\left(\frac{\nB(\omega) + \nF^{e/o}(E)}{s^2+(E-\omega)^2} + \frac{1+\nB(\omega) - \nF^{e/o}(E)}{s^2+(E+\omega)^2} \right).
\end{align}
Keeping track of the differences between $g^{e}$ and $g^o$ leads to modifications of Eqs.~\eqref{disollaplace} and \eqref{rhoMdiffstracesol}. We find
\begin{align}
\tilde d_\alpha^{\,e/o}(s) &= \frac{1}{s+4 \tilde g^{\,e/o}(s)}\ d_\alpha^{\,e/o}(t=0),
\label{eqn:CoherenceFiniteTemp}
\\
\tilde p^{\, e/o}(s) &= \frac{p^{e/o}(t=0) + 4 \tilde g^{\,o/e}(s)/s}{s+4[\tilde g^{\,e/o}(s) +\tilde g^{\,o/e}(s)]}.\label{eqn::FiniteTemp}
\end{align}
We now  observe that $\tilde d_\alpha^{\,e/o}(s)$ has a pole at $s=0$, and that the first term within the brackets in Eq.~\eqref{gevenodd} is divergent for $\omega=E$ when $\text{Re}(s) = 0$ and $T>0$. As shown in App.~\ref{App::DecayRate}, this implies $d_\alpha^{\,e/o}(t\to\infty) =0$ for all finite $T$, in accordance with the Markovian results discussed in Sec.~\ref{sec5b}.
For {asymptotically long times},  $t\to \infty$, the decay law follows by expanding $\tilde g^{\,e/o}(s)$ for $s\to 0$, as we show in detail in App.~\ref{App::DecayRate}. All coherences then die out exponentially,
\begin{equation}\label{longtime}
    d_\alpha^{\,e/o}(t) \propto e^{-\Gamma^{e/o} t},
\end{equation}
where we obtain the same decay rates $\Gamma^{e/o}$ as from the Markovian approach, see Eq.~\eqref{disscons}.
As expected intuitively, environmental memory effects are thus erased at very long times.

Finally, we discuss the long-time behavior of $p^{\,e/o}(t)$ which illustrates the equilibration of the system. Again the result follows by expanding $\tilde p^{\,e/o}(s)$
in Eq.~\eqref{eqn::FiniteTemp} for small $s$, see App.~\ref{App::DecayRate} for details. We find that at $T=0$, Eq.~\eqref{rhoMdiffstracesollongtimes} is recovered.
However, for $T>0$, we get
\begin{eqnarray}\nonumber
p^{\,e/o}(t\to \infty) &=& \frac{\int_\Delta^\infty dE f^{o/e}(E)}{\int_\Delta^\infty dE\left(f^{e/o}(E) +f^{o/e}(E)\right)} \\ &=& \frac{\Gamma^{o/e}}{\Gamma^{e/o}+\Gamma^{o/e}},\label{eqn::tinftyFiniteTemp}
\end{eqnarray}
with the function $f^{e/o}(E)$ in Eq.~\eqref{feo}.
Equation \eqref{eqn::tinftyFiniteTemp} also matches the corresponding result in the Markovian limit, see Eq.~\eqref{peq}.

\subsection{Case study}\label{sec6}

\begin{figure}
\centering
\includegraphics[width=0.45\textwidth]{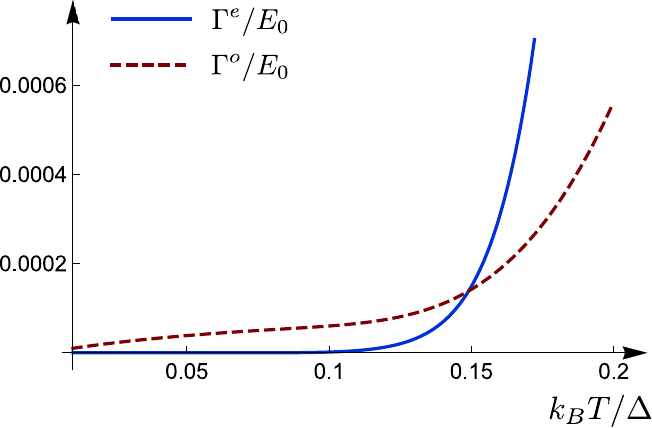}
\caption{Decay rates $\Gamma^{e/o}$ vs temperature $T$ obtained from Eq.~\eqref{disscons} for the spinless $p$-wave TS wire model in Eq.~\eqref{Hamp-wave}.
We use $\omega_0=\Delta$, $\Delta/(mv^2)=0.2$,  $d_SV_S\Delta=850$, and $E_0=e^2/C_0$, see Eq.~\eqref{spectralfunction}. }\label{fig5}
\end{figure}

Here, we provide concrete estimates to illustrate the above results for a specific TS nanowire model. To that end, we use a spinless model for a TS wire with $p$-wave pairing symmetry.  One can write the corresponding BdG Hamiltonian in the form  \cite{AliceaReview}
\begin{equation}\label{Hamp-wave}
{\cal H}_{\rm BdG} = \frac{p^2}{2m} \tau_z  - \Delta \tau_z + vp \tau_x.
\end{equation}
We focus on a semi-infinite wire in order to obtain the zero-energy MBS wave function, $\ket{\text{MBS}}$, as well as the above-gap quasi-particle wave functions, $\ket{k}$.
Given these wave functions, we then compute the $W$ matrix elements needed in Eqs.~\eqref{Wkirdef} and \eqref{widef}. The result can be found in App.~\ref{App::WandJ}, where Fig.~\ref{fig10}  shows a plot of $\nu(E)|W(E)|^2$. In order to evaluate $\delta F$ from Eqs.~\eqref{Neffdif} and \eqref{Neff2},
we assume the dimensionless parameter $d_S V_S \Delta=850$. To obtain this value, we employed the Fermi energy for Al ($11.7$~eV) and the volume $V_S$ as for the experimental setup in Ref.~\cite{Albrecht2016}. The nanowires in the latter experiment were fairly short, but since we are interested in describing the states at just one nanowire end, such a reduced volume should be appropriate. We use the gap value for Al, $\Delta = 2\times 10^{-4}$~eV, and throughout focus on the topological parameter regime, $\Delta>0$.  For simplicity, we will consider the case of relatively small TS gap, $\Delta/(mv^2)\le 1/2$, since the solution described in App.~\ref{App::WandJ} otherwise becomes slightly more involved.
Finally, the electromagnetic environment is fully characterized by specifying the frequency $\omega_0$ and the energy scale $E_0=e^2/C_0$, see Eq.~\eqref{spectralfunction}.

\subsubsection{Finite-$T$ decay rates}
\label{sec6a}

In Fig.~\ref{fig5}, we show the temperature dependence of the decay rates $\Gamma^{e/o}$, Eq.~\eqref{disscons}, when using the BdG Hamiltonian in Eq.~\eqref{Hamp-wave}.
For $\kB T < 0.1\Delta$, we observe that $\Gamma^{e}(T)$ remains exponentially small, in contrast to what is found for the rate $\Gamma^o$ in the odd parity sector.
We thus expect that in this low-temperature regime, the $T=0$ results presented in Sec.~\ref{sec5c} should also apply for the even parity sector at intermediate times.
In particular, for long times but subject to the condition $t\ll\Gamma^{-1}$, where $\Gamma=(\Gamma^e+\Gamma^o)/2$, the off-diagonal entries of $\rrho_M^e(t)$ are expected to remain approximately constant, $d^{\,e}_\alpha(t)\simeq R^e_\alpha$ (with $\alpha=x,y,z$).  Neglecting the effects of early-time transients,
$R^e_\alpha$ is given by the residue of $\tilde d_\alpha^{\,e}(s)$, Eq.~\eqref{eqn:CoherenceFiniteTemp}, at the pole $s=-\Gamma^{e}$.
Keeping for the moment both parity sectors, we have
\begin{align}
R_{\alpha}^{e/o} &= \lim_{s\to -\Gamma^{e/o}} \frac{s+\Gamma^{e/o}}{s+4 \tilde g^{\,e/o}(s)} \, d_\alpha^{\,e/o}(t=0).\label{eqn::ResidueGeneral}
\end{align}
Using the fact that $\Gamma^{e/o}\ll \Delta$, Eq.~\eqref{eqn::ResidueGeneral} can be simplified to
\begin{equation}\label{RRR}
R_{\alpha}^{e/o} = \frac{d_\alpha^{\,e/o}(t=0)}{1+\zeta^{e/o}(T)},
\end{equation}
with
\begin{eqnarray}\label{zeta}
&& \zeta^{e/o} (T)= \frac{4}{\pi} \int_\Delta^\infty dE \int_0^\infty  d\omega \, \nu(E) \abs{W(E)}^2 J(\omega) \\ \nonumber &&\quad \times \left( \frac{1+\nB(\omega) - \nF^{e/o}(E)}{(E+\omega)^2}+ \frac{\nB(\omega)+\nF^{e/o}(E)}{(\Gamma^{e/o})^2+(E-\omega)^2} \right) .
\end{eqnarray}
Noting that $\zeta^{e/o}(T=0)=\eta$, see Eq.~\eqref{alphadef}, we first confirm that Eq.~\eqref{RRR} correctly
recovers the $T=0$ result  \eqref{disollaplacelongtimes}. For finite but low $T$ and focusing on the even parity sector,
the coherence reduction saturates at the value $R_{\alpha}^e$ in Eq.~\eqref{RRR} for intermediate-to-long times, $\Delta^{-1}\ll t<\Gamma^{-1}$.
However, for $t>\Gamma^{-1}$, all coherences will ultimately decay to zero.

\begin{figure}
\centering
\includegraphics[width=0.45\textwidth]{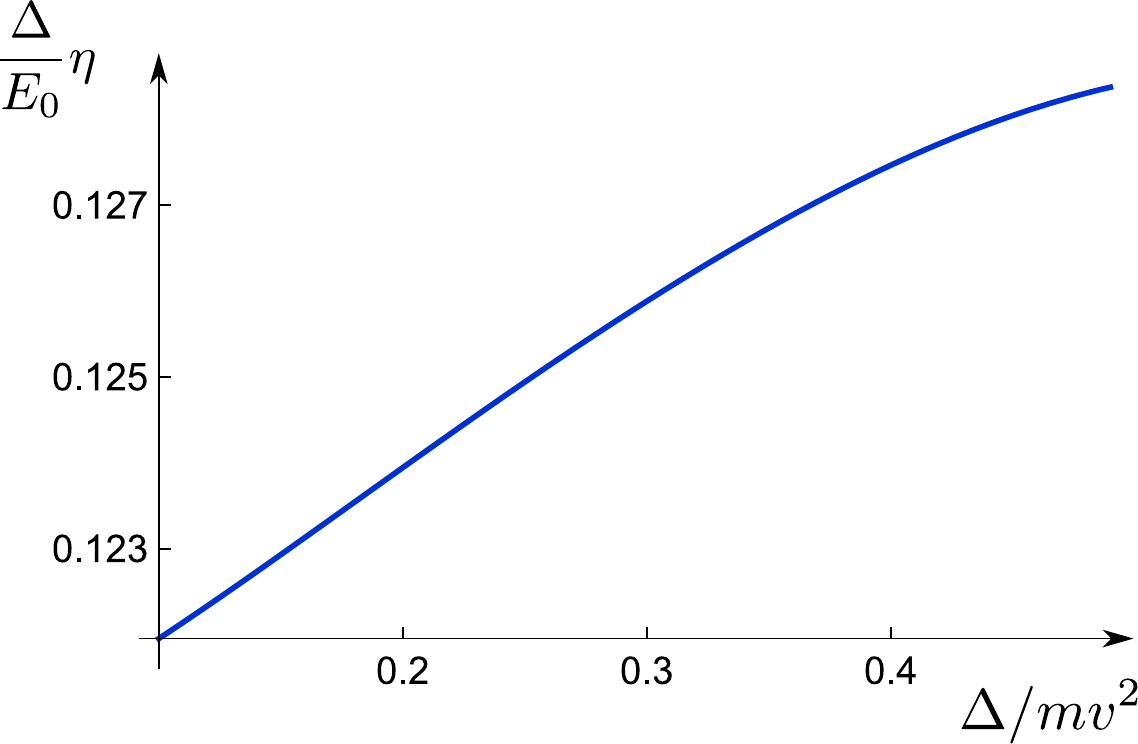}
\caption{ Decoherence parameter $\eta$, see Eq.~\eqref{alphadef}, vs $\Delta/(mv^2)$
for the $p$-wave TS nanowire model in Eq.~\eqref{Hamp-wave}.  We assume an environmental frequency
$\omega_0=\Delta$, other parameters are described in the text.
All coherences are reduced by a factor $1/(1+\eta)$ at long times.
Since we have rescaled $\eta$ by $E_0/\Delta$ in the plot,
the shown results hold for arbitrary ratio $E_0/\Delta$.}\label{fig6}
\end{figure}

\begin{figure}
\centering
\includegraphics[width=0.45\textwidth]{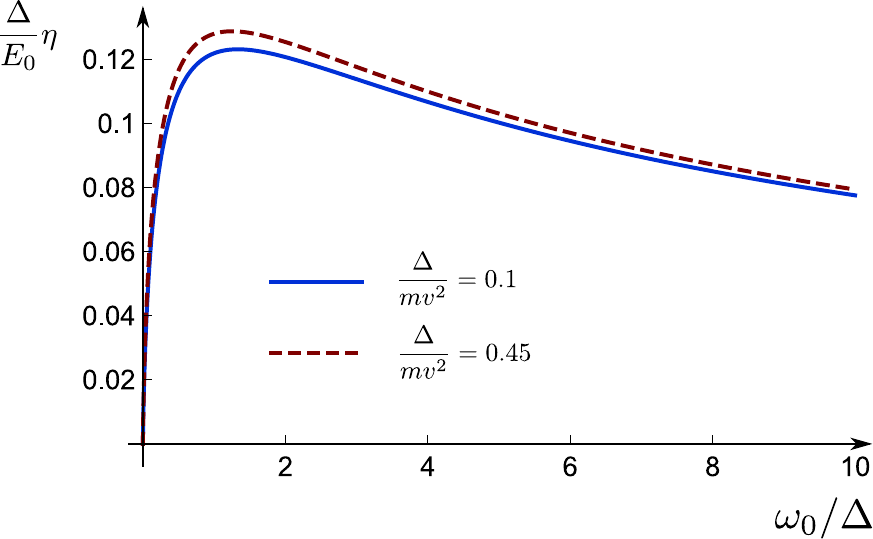}
\caption{Decoherence parameter $\eta$ vs $\omega_0/\Delta$ for $\Delta/(mv^2)=0.1$ and $\Delta/(mv^2)=0.45$, cf.~the caption of Fig.~\ref{fig6}.  }\label{fig7}
\end{figure}

\subsubsection{Zero-temperature fidelity reduction}
\label{sec6b}

We found in Sec.~\ref{sec5c} that even at zero temperature, quantum fluctuations in the electrodynamic environment can generate virtual (off-shell) processes that, on the non-Markovian level, cause a fidelity reduction in the readout of the bare-Majorana qubit.  The efficiency of this process is encoded by the dimensionless parameter $\eta$ in Eq.~\eqref{alphadef},
where all long-time coherences, $d_\alpha^{\, e/o}(t\to \infty)$, are reduced by a common factor $1/(1+\eta)$ with respect to their initial value, see Eq.~\eqref{disollaplacelongtimes} and the qualitative discussion in Sec.~\ref{sec4}. In Figs.~\ref{fig6} and \ref{fig7}, we show the dependence of $\eta$ on the dimensionless parameters $\Delta/(mv^2)$ and $\omega_0/\Delta$, respectively.
Since $\eta$ has been rescaled by $E_0/\Delta$ in both figures, these results are valid for arbitrary $E_0/\Delta$. In fact, for large values of $E_0/\Delta$ one gets
large values of $\eta$ and hence a strong suppression of the coherences. To minimize the reduction, one should thus minimize $E_0=e^2/C_0$.

Apart from its significance for quantum information processing applications, the $T=0$ fidelity reduction for bare-Majorana qubits is also of importance from a theoretical point of view.
Figure \ref{fig7} indicates that this effect is most pronounced for $\omega_0\approx \Delta$, where quantum fluctuations of the Ohmic electromagnetic environment can almost resonantly match the TS gap. In addition, Fig.~\ref{fig6} shows that $\eta$ grows with decreasing TS gap.
This can be rationalized by noting that the Ohmic spectral function \eqref{spectralfunction} includes gapless low-energy bosons  that can participate in the coherence reduction, see Sec.~\ref{sec4}.

\begin{figure}
\centering
\includegraphics[width=0.47\textwidth]{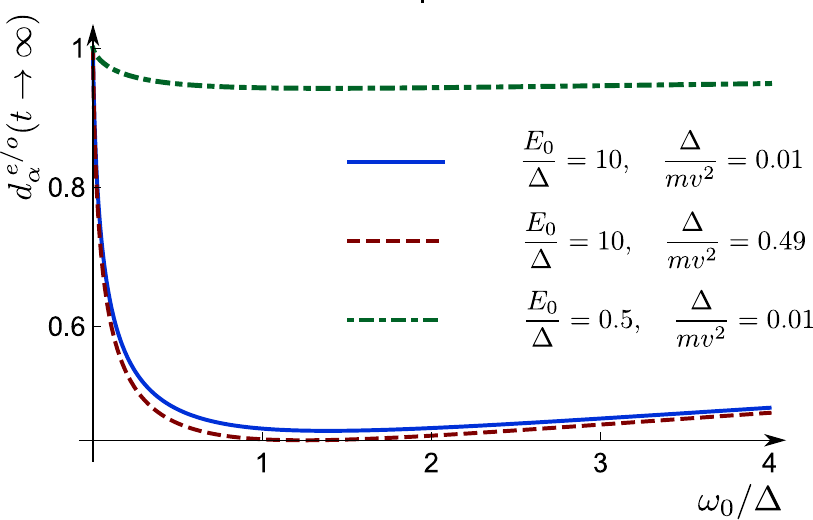}
\caption{Asymptotic $T=0$ long-time coherences, $d^{\,e/o}_\alpha(t\to \infty)$ [in units of $d^{\,e/o}_\alpha(0)$], vs  $\omega_0/\Delta$.  Results are shown for several parameter sets ($E_0/\Delta, \Delta/mv^2$), cf.~Eq.~\eqref{disollaplacelongtimes}, and neither depend on  $\alpha(=x,y,z)$ nor on the parity ($e/o$) index.
 }\label{fig8}
\end{figure}

\begin{figure}
\centering
\includegraphics[width=0.47\textwidth]{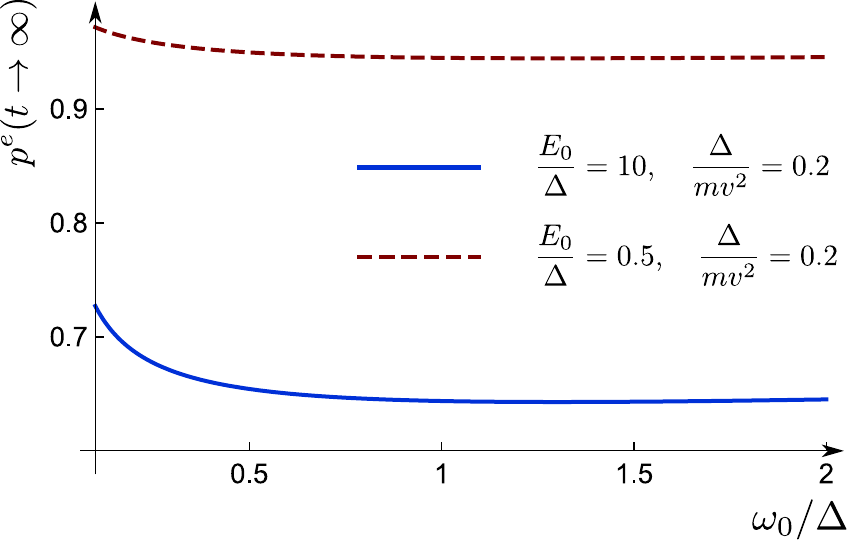}
\caption{Long-time $T=0$ probability for staying in the even parity sector, $p^{e}(t\to\infty)$ in Eq.~\eqref{rhoMdiffstracesollongtimes}, vs  $\omega_0/\Delta$, for $p^e(0) =1$, $\Delta/mv^2=0.2$, and two different values for $E_0/\Delta$. }\label{fig9}
\end{figure}
In Fig.~\ref{fig8}, we illustrate the value of $d_\alpha^{\,e/o}(t)$ reached at long times in the $T=0$ limit.  We observe that especially for large $E_0/\Delta$ and $\omega_0\approx \Delta$, the coherence reduction is quite significant.
Finally, Fig.~\ref{fig9} depicts the $\omega_0/\Delta$-dependence of the $T=0$ probability for staying in the even parity Majorana sector at very long times, $p^e(t\to \infty)$, provided that one has started out from this sector, $p^e(0)=1$.  The analytical prediction for this quantity is given by $(1+\eta)/(1+2\eta)\ge 1/2$, see Eq.~\eqref{rhoMdiffstracesollongtimes}.
We find that for large $E_0/\Delta$, the parity reduction can be rather large.
Taking, say, $E_0/\Delta=10$ and $\omega_0\approx \Delta$,
a parity leakage of $\approx 35$ from the even into the odd parity Majorana sector is observed in Fig.~\ref{fig9}.

\section{Conclusions}\label{sec7}

We have re-examined the issue of decoherence of qubits formed by zero-energy Majorana bound states when coupled to an electromagnetic environment that causes transition matrix elements between the qubit and the above-gap states. The environment is described by a Caldeira-Leggett bath of non-interacting bosons with an Ohmic spectral density \cite{Weiss2010}. Concrete estimates have been provided in Sec.~\ref{sec6} for a specific microscopic superconductor model, where the topological superconductor corresponds to a spinless nanowire with $p$-wave pairing.

We have pointed that if the MBSs do not overlap, there is still in principle full topological protection, but the parity is not shared between the MBS and the quasiparticle continuum. Therefore in order to take advantage of the protection it is necessary that the readout couples to the dressed states, \textit{i.e.}, the MBS dressed by bosons and continuum quasiparticles. Related proposals for the operation of topological qubits in this basis were discussed in Refs. \onlinecite{Aasen2016} and \onlinecite{Akhmerov2010}, and we have here pointed out that there are limitations when using quantum-dot readout because of lack of adiabaticity. The time scale of switching on the quantum dot will be extremely important for the fidelity of the readout.

We have studied in detail the situation for a projective measurement of the bare (undressed) MBS. Our theoretical approach is based on a modified Bloch-Redfield quantum master equation for the reduced density matrix of the bare-Majorana qubit, and it holds for weak coupling between the Majorana sector and the environment. In formulating this theory, we have carefully accounted for the fact that total fermion number parity is conserved (within our model) and we have emphasized that it is necessary to keep track of the entanglement between the Majorana subsystem and environmental degrees of freedom. For a quantitative description, the virtual off-shell scattering processes behind this physics require a full non-Markovian master equation approach. From this approach, we find that the off-diagonal elements of the reduced density matrix of the isolated Majorana subsystem (the \textit{bare-Majorana qubit}), taken at $T=0$, become suppressed by a factor $1/(1+\eta)$ at long times, where $\eta$ is defined in Eq.~\eqref{alphadef}.  The fidelity therefore saturates at a reduced but finite value at $T=0$. On a qualitative level, this conclusion already follows from a simple perturbative consideration, see Sec.~\ref{sec4}. Likewise, the probability to remain in a given parity sector of the Majorana subsystem will be reduced by a finite amount. With minor modifications, our $T=0$ results also describe the case of very low but finite temperatures when considering the decoherence dynamics on intermediate-to-long time scales, $\Delta^{-1}\ll t<\Gamma^{-1}$, see Sec.~\ref{sec6a}. At finite temperatures,  the asymptotic long-time behavior of the decoherence dynamics is well described by the Markovian approximation which has also  been used in most previous theories \cite{Goldstein2011,Cheng2012,Rainis2012,Schmidt2012,Yang2014,Lai2018,Song2018,Li2018,Nag2018,Aseev2018}.

The important fidelity-reduction parameter $\eta$ in Eq.~\eqref{alphadef} depends on the spectral density of the electromagnetic environment, on the quasi-particle density of states, and on a function $W(E)$ which encodes the transition matrix elements between Majorana and quasi-particle states. Physical conditions for when $\eta$  becomes significant have been specified in detail in Sec.~\ref{sec6}.

We conclude by noting that fluctuating gate charges are ubiquitous in candidate devices for realizing Majorana qubits. For that reason, the fidelity reduction discussed in this paper may constitute an important limitation for the coherent operation of Majorana qubits. However, our theory also shows the fidelity-reduction parameter $\eta$ could be minimized by proper parameter choices and we point out that it would be extremely interesting for future studies to determine how one can minimize the fidelity-reduction by careful timing of the readout protocol.

\begin{acknowledgments}
We thank Torsten Karzig, Christina Knapp, Chetan Nayak, Yuval Oreg, Mark Rudner, and Ady Stern for helpful discussions. We acknowledge support by the Danish National Research Foundation as well as by  the Deutsche Forschungsgemeinschaft (DFG, German Research Foundation) –
Projektnummer 277101999 – TRR 183 (project C01).
\end{acknowledgments}

\appendix
\section{On the finite-$T$ non-Markovian case}\label{App::DecayRate}

We here provide additional details concerning Sec.~\ref{sec5d}.
We first give a detailed derivation of Eq.~\eqref{longtime} describing  the long-time dephasing dynamics.
In general, the long-time limit is dominated by small-$s$ contributions in the Laplace transformed picture.
We start by examining the small-$s$ form of the Laplace transformed functions $\tilde g^{\,e/o}(s)$ in Eq.~\eqref{gevenodd}.
To lowest order in $s$, the second term of Eq.~\eqref{gevenodd} equals $2sA^{e/o}$ with
\begin{eqnarray} \nonumber
A^{e/o} &=& \int_0^\infty\frac{d\omega}{2\pi}\int_\Delta^\infty dE\,\frac{J(\omega)\nu(E)|W(E)|^2}{(\omega+E)^2} \\
 &\times& \left[ 1+\nB(\omega) - \nF^{e/o}(E)  \right].
\end{eqnarray}
For the first term of Eq.~\eqref{gevenodd}, we change variables to $\omega_\pm=(\omega\pm E)/2$, with integral limits  $\omega_+\in [\Delta/2,\infty)$ and $\omega_-\in [-\omega_+, \omega_+-\Delta]$.
For $s=0$, the integrand in Eq.~\eqref{gevenodd} diverges as $\omega_-\to 0$. This divergence happens outside the integration limits when  $\omega_+<\Delta$.  The contribution from $\omega_+\in[\Delta/2,\Delta)$ can thus safely be evaluated by putting $s=0$ in the integrand.  The result is written as $sK^{e/o}/4$ with
\begin{eqnarray}
K^{e/o} &=& \frac{2}{\pi}\int_{\Delta/2}^\Delta d\omega_+ \int_{-\omega_+}^{\omega_+-\Delta} \frac{ d\omega_-}{\omega_-^2} \nu(\omega_+-\omega_-)
\nonumber \\ &\times & \abs{W(\omega_+-\omega_-)}^2 J(\omega_++\omega_-) \\ \nonumber &\times& \left[\nB(\omega_++\omega_-)+\nF^{e/o}(\omega_+-\omega_-)\right].
\end{eqnarray}
In the remaining part of $\tilde g^{\,e/o}(s)$, the dominant contribution from the $\omega_-$ integral is picked up around $\omega_-=0$, and so we approximate the integrand by evaluating all terms except for the $1/(s^2+\omega_-^2)$ factor at $\omega_-=0$. With $f^{e/o}(\omega)$ in Eq.~\eqref{feo},
this results in a third contribution to $\tilde g^{\,e/o}(s)$ of the form
\[
\frac{s}{2\pi} \int_{\Delta}^\infty d\omega_+ \;f^{e/o}(\omega_+) \int_{-\omega_+}^{\omega_+-\Delta}  \frac{d\omega_-}{s^2+4\omega_-^2}.
\]
Performing the $\omega_-$-integration, renaming $\omega_+\to E$, and collecting all terms,  we arrive at the small-$s$ expansion
\begin{eqnarray}\label{eqn::geoFiniteTempSeriesins}
\tilde g^{\,e/o}(s) &=& 2sA^{e/o}+\frac{sK^{e/o}}{4}+ \frac{1}{4} \int_{\Delta}^\infty dE f^{e/o}(E)\\
\nonumber
&-&\frac{s}{8\pi} \int_{\Delta}^\infty dE f^{e/o}(E)  \frac{2E-\Delta}{E(E-\Delta)} +{\cal O}(s^2).
\end{eqnarray}

From Eq.~\eqref{eqn:CoherenceFiniteTemp}, we then find a pole for the Laplace transform of the coherences, $\tilde d_\alpha^{\,e/o}(s)$. This pole dictates the long-time behavior of $d_\alpha^{\,e/o}(t)$.
For $t\to \infty$, we thereby arrive at Eq.~\eqref{longtime} where the rates are given by
\begin{eqnarray}\nonumber
\Gamma^{e/o}& =& \frac{M\int_{\Delta}^\infty dE f^{e/o}(E)}{1 + 8 A^{e/o}+K^{e/o}-\int_{\Delta}^\infty \frac{dE}{2\pi} f^{e/o}(E)  \frac{2E-\Delta}{E(E-\Delta)}}
\\  &\simeq& M\int_\Delta^\infty  dE f^{e/o}(E).
\label{Eqn::app::Omega}
\end{eqnarray}
In the last step, we have used that the coupling between the Majorana system and the environment is weak.
 The final result for these  rates coincides with the corresponding Markovian result \eqref{disscons}.

Next we address the asymptotic values  $p^{\,e/o}(t\to\infty)$, which  follow by inserting the small-$s$ expansion of $\tilde g^{\,e/o}(s)$ in Eq.~\eqref{eqn::geoFiniteTempSeriesins} into Eq.~\eqref{eqn::FiniteTemp}. We then find that $\tilde p^{\,e/o}(s)$ has a pole at $s=0$. At finite $T$, only the $s$-independent term in Eq.~\eqref{eqn::geoFiniteTempSeriesins} contributes to the residue of $\tilde p^{\,e/o}(s)$ at $s=0$, and thus Eq.~\eqref{eqn::tinftyFiniteTemp} follows. On the other hand, the $T=0$ result for $p^{\,e/o}(t\to\infty)$, see Eq.~\eqref{rhoMdiffstracesollongtimes}, is recovered by noting that the only non-vanishing $T=0$ term in Eq.~\eqref{eqn::geoFiniteTempSeriesins} comes from $A^{e/o}$.
Some algebra then leads to Eq.~\eqref{rhoMdiffstracesollongtimes}.

\section{Solution of TS nanowire model}\label{App::WandJ}

In what follows, we discuss the solution of the specific TS nanowire model in Eq.~\eqref{Hamp-wave} and determine the $W$ matrix elements which encode the energy-dependent transition matrix elements between the MBS subsystem and the quasi-particle sector.
These results have been used for generating the numerical data shown in Secs.~\ref{sec5} and \ref{sec6}.
We consider the BdG Hamiltonian \eqref{Hamp-wave} for a spinless semi-infinite TS nanowire with 1D coordinate $x\ge 0$. We first write Eq.~\eqref{Hamp-wave} in the equivalent form
\begin{equation}
    {\cal H}_{\rm BdG}=\Delta \left[\left(\frac{\tilde{p}^2}{2\delta} - 1   \right)\tau_z + \frac{\tilde{p}}{\delta}\tau_x \right],
\end{equation}
where we define
\begin{equation}
    \tilde{p}=\frac{p}{mv},\quad \delta=\frac{\Delta}{mv^2}.
    \end{equation}
Similarly, we use the notation $\tilde{k} =k/(mv)$ below.
The zero-energy MBS wave function is denoted by $\psi_0(x)=\braket{x}{\text{MBS}}$, and quasi-particle wave functions  by $\psi_k(x)=\braket{x}{k}$.
 With the \emph{Ansatz} $\psi_0(x) = \chi_0 e^{ik_0x}$, normalizable MBS solutions are found for $k_0$ with positive imaginary values, $k_0=i\kappa_0^\pm$, where
\begin{equation}\label{App::k0}
    \kappa_{0}^\pm=mv(1\pm \sqrt{1-2\delta}).
\end{equation}
We only consider the regime $0\leq \delta\leq 1/2$ here and in Sec.~\ref{sec6}.

\begin{figure}[t]
\centering
\includegraphics[width=0.47\textwidth]{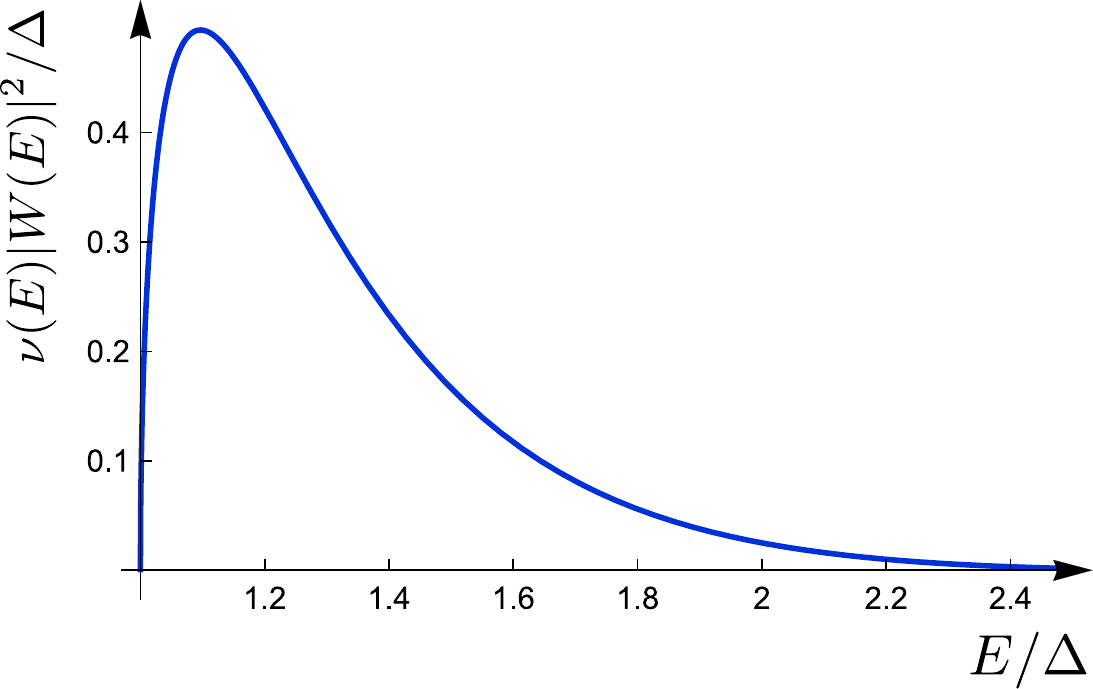}
\caption{Transition matrix element $\nu(E)\abs{W(E)}^2$ vs energy $E$ for the TS nanowire model \eqref{Hamp-wave} with $\Delta/(mv^2)=0.2$.
 }\label{fig10}
\end{figure}

Taking a linear superposition of the two states corresponding to Eq.~\eqref{App::k0}, and imposing Dirichlet boundary conditions, $\psi_0(0)=0$, we obtain the Nambu spinor wave function for the MBS,
\begin{equation}
    \psi_0(x) = \frac{1}{\mathcal{N}_0} \left( e^{-\kappa_{0}^+ x }- e^{-\kappa_{0}^- x  } \right)\begin{pmatrix} 1\\ -i \end{pmatrix},\quad \mathcal{N}_0 = \sqrt{\frac{1-2\delta}{mv\delta}}.
\end{equation}
As expected, this wave function is exponentially localized near the boundary at $x=0$. Similarly,
quasi-particle wave functions follow from the \emph{Ansatz} $\psi_k(x)= \chi_k e^{ik x}$, with $E_k\ge \Delta$ given by
\begin{equation}
    E_k = \Delta  \left[\left(\frac{\tilde{k}^2}{2\delta} -1 \right)^2 +\frac{\tilde{k}^2}{\delta^2} \right]^{1/2} .
\end{equation}
We then find four solutions, $k=\pm k_s$  (with $s=\pm$),
\begin{align}
    k_s =  \sqrt2 mv \sqrt{\delta-1+ s \sqrt{1-2\delta  + \delta^2 \xi^2}},\label{AppB::k}
\end{align}
with $\xi_k = E_k/\Delta$.
For  $0\leq \delta \leq 1/2$, we observe that $k_-=i\kappa$ (with $\kappa>0$) is purely imaginary while $k_+$ is purely real.
Dropping the non-normalizable states with $k=-i\kappa$, we write $k_+=k$.
We now impose Dirichlet boundary conditions at $x=0$. Exploiting the continuity equation at large $x$, we find
\begin{eqnarray}\nonumber
    \mathcal{N}_k\psi_k(x) &=&
    \begin{pmatrix}{\tilde{k}}/{\delta} \\a_k \end{pmatrix}
    e^{i kx} + \begin{pmatrix}-{\tilde{k}}/{\delta} \\ a_k \end{pmatrix} e^{-ik x + i\theta_k}\\
    &-&   \varepsilon_k \begin{pmatrix}i{\tilde{\kappa}}/{\delta} \\b_k\end{pmatrix} (1+e^{i\theta_k})e^{-\kappa x },
\end{eqnarray}
with
\begin{subequations}
\begin{eqnarray}
    a_k&=& 1+\xi_k-\tilde{k}^2/(2\delta), \\
    b_k&=&1+\xi_k+\tilde{\kappa}^2/(2\delta),
\end{eqnarray}
\end{subequations}
and
\begin{equation}\label{varepsilonktan}
 \varepsilon_k = \frac{a_k}{b_k} ,\quad \tan\frac{\theta_k}{2} = \frac{\kappa}{k}\varepsilon_k.
\end{equation}
The normalization constant follows from
\begin{equation}\label{N2k}
\mathcal{N}^2_k = 2 L\tilde{k}^2/\delta^2 + 2L a_k^2 .
\end{equation}
Here $L$ is wire length, where we let $L\to \infty$ in the end. Equation \eqref{Wkirdef} then yields
\begin{widetext}
\begin{equation}\label{Wk}
\begin{aligned}
    W_k =&  \frac{ 4i e^{-i{\theta_k}/{2}} }{\tilde{\mathcal{N}}\sqrt{mvL}} \Bigg[\frac{1}{(2\delta-\tilde{k}^2)^2+4\tilde{k}^2}
     \bigg(\left(2\tilde{k}^2/\delta - a_k (2\delta-\tilde{k}^2) \right) \cos\left(\frac{\theta_k}{2}\right)
     -\left( 2a_k \tilde{k}+2\tilde{k}-\tilde{k}^3/\delta  \right) \sin\left(\frac{\theta_k}{2} \right) \bigg)\\
     &\qquad -\varepsilon_k \cos\left(\frac{\theta_k}{2}\right) \frac{\tilde\kappa/\delta -b_k}{\tilde{\kappa}^2+2(\tilde{\kappa}+\delta) }\Bigg],
\end{aligned}
\end{equation}
\end{widetext}
where
\begin{equation}
     \tilde{\mathcal{N}}= \sqrt{\frac{2}{\delta} \left(\frac{\tilde{k}^2}{\delta^2} + a_k^2 \right)}.
\end{equation}

Finally, $\nu(E)\abs{W(E)}^2$ follows from Eq.~\eqref{widef} by observing that the density of states is with $k=k_+(E)$ in Eq.~\eqref{AppB::k} given by
\begin{equation}
    \nu(E)=\sum_k \delta(E-E_k) = \frac{L}{2\pi} \frac{dk}{dE}.
    \end{equation}
Note that the $L$-dependent prefactors in $\abs{W(E)}^2$ are cancelled by those in $\nu(E)$.  Figure \ref{fig10} shows a plot of the resulting
product $\nu(E)\abs{W(E)}^2$.

\section{Approximate Laplace transform}
\label{app:laplace}

We here provide details about the numerical inverse Laplace transformation used for generating Fig.~\ref{fig4}.
We start with the Laplace transformed function $\tilde g^{\,e/o}(s)$ in Eq.~\eqref{gevenodd}, which at $T=0$
becomes parity independent and given by
\begin{align}\label{gevenodd2}
\tilde g_0(s)
= \frac{s}{\pi}\int_0^\infty d\omega \int_\Delta^\infty dE\, \frac{\nu(E) \abs{W(E)}^2 J(\omega)}{s^2+(E+\omega)^2}.
\end{align}
Since $\nu(E) \abs{W(E)}^2$ is peaked at $E=\Delta_p$, where $\Delta_p$ is slightly above $\Delta$, see Fig.~\ref{fig10}, we write
\begin{align}
\tilde g_0(s)&\approx \frac{s}{\pi} \int_0^\infty d\omega\,\frac{J(\omega) }{s^2+(\Delta_p+\omega)^2}\int_\Delta^\infty dE \ \nu(E) \abs{W(E)}^2.
\end{align}
 Inserting $J(\omega)$ from Eq.~\eqref{spectralfunction}, we encounter the auxiliary function
\begin{equation}\label{intw}
 \tilde h_0(s) = 2s\int_0^\infty d\omega\, \frac{\omega}{\omega^2+\omega_0^2}\frac{1}{s^2+(\Delta_p+\omega)^2}.
\end{equation}
For Re$(s)>0$, this yields
\begin{widetext}
\begin{equation}\label{hres}
  \tilde h_0(s)=\omega_0\frac{-\pi\Delta_p[\Delta_p^2+(s+\omega_0)^2]+2\Delta_p [\Delta_p^2+s^2+\omega_0^2]\tan^{-1}(\Delta_p/s)+s(\Delta^2_p-\omega_0^2+s^2)\ln\left[(s^2+\Delta_p^2)/\omega_0^2\right]}
  {[\Delta_p^2+(s-\omega_0)^2][\Delta_p^2+(s+\omega_0)^2]}.
\end{equation}
For the Laplace transformed coherences in Eq.~\eqref{disollaplace}, we then obtain
\begin{equation}\label{disollaplacehs}
 \tilde d_\alpha^{\,e/o}(s)= \frac{d_\alpha^{\,e/o}(t=0) }{s+B \tilde h_0(s)},\quad B=\frac{E_0}{\pi} \int_\Delta^\infty dE \ \nu(E) \abs{W(E)}^2.
\end{equation}
At this stage, the inverse Laplace transform can be performed numerically in an efficient manner, see  Fig.~\ref{fig4}.

For finite but very low temperatures, $\kB T\ll \Delta$, we should keep the Bose function $\nB(\omega)$ in Eq.~\eqref{gevenodd}. The function $\tilde h_0(s)$ should then be replaced by $\tilde h(s)=\tilde h_0(s)+\tilde h_1(s)$, where
\begin{equation}\label{intwT}
 \tilde h_1(s)=  2s\int_0^\infty d\omega\, \frac{\omega\nB(\omega)}{\omega^2+\omega_0^2}\left(\frac{1}{s^2+(\Delta_p+\omega)^2}+\frac{1}{s^2+(\Delta_p-\omega)^2}\right).
\end{equation}
\end{widetext}
We see that the saturation value $d_\alpha^{\,e/o}(t\to \infty)$, which follows by setting $s=0$, now vanishes because $\tilde h_1(0)$ diverges. This feature is a general result of the exponential decay of all coherences in the Markovian case with $T>0$.
Finally, we remark that finite temperature also gives only minor modifications to the dynamics shown in Fig.~\ref{fig4}.

\end{document}